\documentclass[journal,10pt]{IEEEtran}

\usepackage[T1]{fontenc}
\usepackage{amsfonts}
\usepackage{amsmath}
\usepackage{amsthm}
\usepackage{amssymb}
\usepackage{caption}
\usepackage[labelformat=simple]{subcaption}
\usepackage{color}
\usepackage{cite,graphicx,algorithm}
\usepackage{algorithmic}
\usepackage{bm}

\ifCLASSINFOpdf
\else
\fi
\theoremstyle{plain}

\newtheorem{lemma}{Lemma}

\newtheorem{remark}{Remark}

\begin{document}
%
\title{Energy Efficiency for Proactive Eavesdropping in Cooperative Cognitive Radio Networks}
%
%
%

\author{Yao~Ge,~\IEEEmembership{Member,~IEEE,}
       and~P.~C.~Ching,~\IEEEmembership{Fellow,~IEEE}
\thanks{Y. Ge and P. C. Ching are with the Department
of Electronic Engineering, The Chinese University of Hong Kong, Hong Kong SAR of China (e-mail: yaoge.gy.jay@hotmail.com; pcching@ee.cuhk.edu.hk).}}

%
%

\markboth{}%
{Shell \MakeLowercase{\textit{et al.}}: Bare Demo of IEEEtran.cls for IEEE Communications Society Journals}
%



\maketitle

\begin{abstract}
This paper investigates a distant proactive eavesdropping system in cooperative cognitive radio (CR) networks. Specifically, an amplify-and-forward (AF) full-duplex (FD) secondary transmitter assists to relay the received signal from suspicious users to legitimate monitor for wireless information surveillance. In return, the secondary transmitter is granted to share the spectrum belonging to the suspicious users for its own information transmission. To improve the eavesdropping, the transmitted secondary user's signal can also be used as a jamming signal to moderate the data rate of the suspicious link. We consider two cases, i.e., non-negligible processing delay (NNPD) and negligible processing delay (NPD) at secondary transmitter. Our target is to maximize network energy efficiency (NEE) via jointly optimizing the AF relay matrix and precoding vector at the secondary transmitter, as well as the receiver combining vector at monitor, subject to the maximum power constraint at the secondary transmitter and minimum data rate requirement of the secondary user. We also guarantee that the achievable data rate of the eavesdropping link should be no less than that of the suspicious link for efficient surveillance. Due to the non-convexity of the formulated NEE maximization problem, we develop an efficient path-following algorithm and a robust alternating optimization (AO) method as solutions under perfect and imperfect channel state information (CSI) conditions, respectively. We also analyze the convergence and computational complexity of the proposed schemes. Numerical results are provided to validate the effectiveness of our proposed schemes.
\end{abstract}

\begin{IEEEkeywords}
Cooperative cognitive radio, full-duplex, network energy efficiency, proactive eavesdropping, wireless information surveillance.
\end{IEEEkeywords}

%
\IEEEpeerreviewmaketitle

\section{Introduction}
%
%
%
%

With recent advancement of the Internet of Things (IoT) and 5G wireless communications, infrastructure-free or user-controlled communications (e.g., \textit{ad hoc} networks, device-to-device (D2D) and machine-to-machine (M2M) systems) bring new threats on public security, in which malicious users or terrorists can illegally utilize them to commit crimes or even execute terror attacks. As such malicious misuse is launched through infrastructure-free wireless communications, existing information surveillance methods that rely on cellular or Internet infrastructures are difficult to monitor these situations. Compared to conventional physical layer wireless security intended to prevent eavesdropping, a new research angle by considering eavesdropping as a form of legitimate monitoring was presented in \cite{xu2017surveillance}. Therefore, how to exploit effective proactive eavesdropping techniques to monitor and intervene the infrastructure-free suspicious and malicious wireless communications becomes necessary and challenging.

For wireless communication surveillance, a proactive eavesdropping scheme via jamming \cite{xu2016proactive,xu2017proactive,zhong2017multi,cai2017proactive,huang2018robust,feizi2020proactive} was developed to improve eavesdropping performance, where the legitimate full-duplex
(FD) monitor intercepts information from the suspicious pair and
simultaneously sends a jamming signal to moderate the data rate of the suspicious link, so as to guarantee that the suspicious information can be decoded
successfully at the monitor. To further enhance the eavesdropping ability of the monitor, an advanced spoofing technique was proposed in \cite{zeng2016wireless,xu2018transmit,moon2018relay}, in which
a spoofing relay can forward constructive or destructive signals to the suspicious
receiver when the monitor experiences a strong eavesdropping channel or
suffers from a poor eavesdropping channel, respectively. Wireless information surveillance via proactive eavesdropping has also been investigated in numerous scenarios, such as relay-based suspicious links \cite{ma2017wireless,jiang2017proactive,hu2017proactive,hu2020proactive}, relay-based monitoring links \cite{moon2018proactive,li2018cooperative}, multiple suspicious communication links \cite{li2018wireless,han2019jamming,li2018energy,li2019energy}, and millimeter wave (mmWave) communication systems \cite{cai2018joint}.

Although proactive eavesdropping has gained significant attention from the
research community, there is a paucity of literature focused on cognitive radio (CR) networks. CR has emerged as a promising technology to overcome spectrum scarcity in IoT systems, and can improve spectrum utilization efficiency by allowing spectrum sharing between multiple wireless networks. As the suspicious users may pretend to be legitimate users (e.g., primary users (PUs) or secondary users (SUs)) in CR networks, the investigation of wireless information surveillance schemes in such settings is highly meaningful and important.
Recently, \cite{li2018proactive,xu2020legitimate} considers a CR-based suspicious communication system, in which the suspicious users work as SUs and a FD monitor attempts to intercept the suspicious information via jamming. Furthermore, a novel spectrum sharing incentive mechanism is proposed in \cite{xu2021spectrum} for legitimate wireless information surveillance. Here, the SUs are allowed to access the spectrum of the suspicious users on the condition that they can assist to monitor and contribute to jamming for improved information surveillance.

In this paper, we consider a scenario in which suspicious users are located far away from the monitor M, which makes proactive eavesdropping more difficulty and challenging. To tackle this, we investigate a cooperative CR networks to facilitate distant eavesdropping, where a secondary transmitter T operates in an amplify-and-forward (AF) FD mode, and simultaneously receives information from suspicious users and forwards this eavesdropped signal to the monitor M. In return, authorized agencies allow T to transmit its own message by sharing the spectrum belonging to the suspicious users. It is worth noting that the transmitted SU's signal can also be used as a jamming signal to moderate the data rate of the suspicious link for effective eavesdropping. Therefore, such spectrum sharing scheme can not only help to improve the eavesdropping performance of the monitor, but also provide spectrum access opportunity to the SU.
We consider two cases, namely non-negligible processing delay (NNPD) and negligible processing delay (NPD) at T in our model. Note that energy efficiency (EE) is a critical issue for the design of future IoT communication systems under the background of economic and operational considerations, as well as environmentally-friendly transmission behaviours. Our objective here is to maximize network energy efficiency (NEE) when the available channel state information (CSI) is perfect. We also consider imperfect CSI related to the suspicious link to maximize the worst-case NEE.
The contributions of our work are summarized as follows:
\begin{enumerate}
\item We present a cooperative CR networks to facilitate distant proactive eavesdropping, in which the SUs help the monitor to eavesdrop on the suspicious communication, and meanwhile, transmit its own information, which can also be used as a jamming signal to moderate the data rate of the suspicious link for effective eavesdropping. We consider both the NNPD and NPD cases at T, and compare their performance.
\item We investigate the NEE maximization problem via jointly optimizing the AF relay matrix and the precoding vector at T, as well as the receiver combining vector at the legitimate monitor M, subject to the maximum power constraint at T and the minimum data rate requirement of the SU. We also guarantee that the achievable data rate of the eavesdropping link should be no less than that of the suspicious link for efficient surveillance.
\item We propose an efficient path-following algorithm to address the non-convexity of the formulated NEE maximization problem for perfect CSI. We also develop a robust alternating optimization (AO) method to tackle the formulated worst-case NEE maximization problem for imperfect CSI. We analyze convergence and computational complexity for both schemes.
\item We present numerical results to demonstrate effectiveness of
our proposed schemes. The achieved NEE performance of NPD outperforms that achieved by NNPD for both perfect and imperfect CSIs.
\end{enumerate}

The remainder of this paper is organized as follows: Section \ref{II_model}
describes the system model of cooperative CR networks for proactive eavesdropping. The NEE maximization design for perfect and imperfect CSIs is presented in Section \ref{Perfect_CSI} and Section \ref{IMPerfect_CSI}, respectively. Section \ref{numer} provides numerical results under different scenarios. Our conclusions are finally drawn in Section \ref{conclu}. The Appendix contains
some detailed proofs at the end of the paper.



\section{System Model}\label{II_model}
\begin{figure}
  \centering
  \includegraphics[width=3.2in]{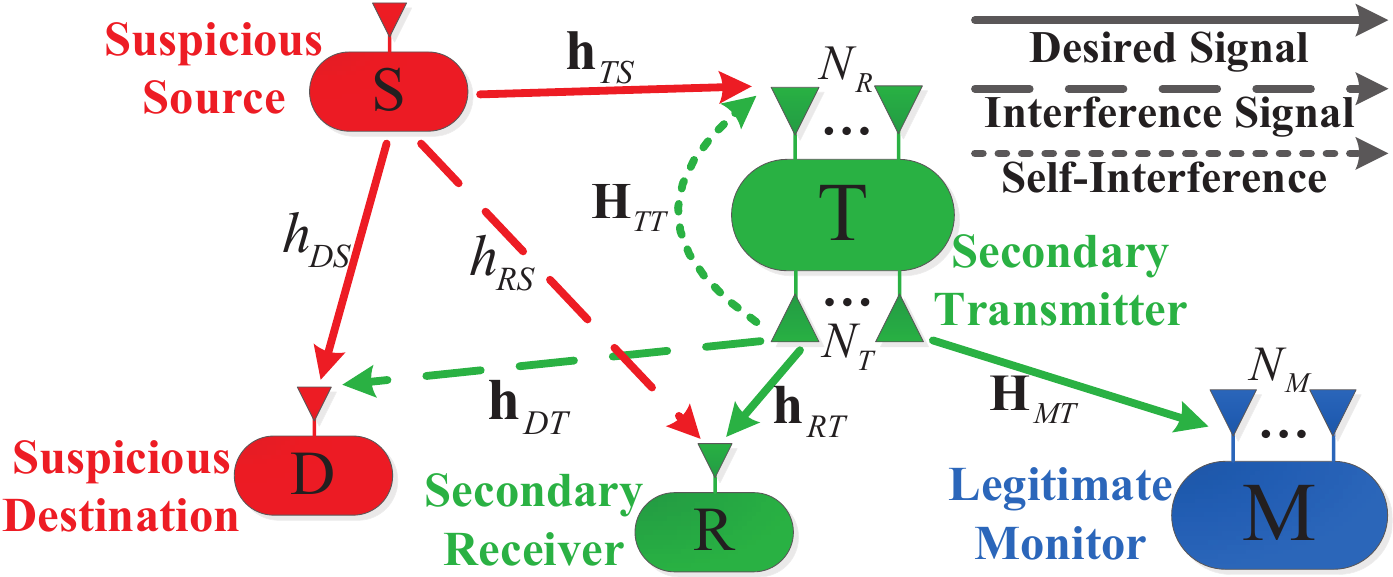}\\
  \caption{Proactive eavesdropping in cooperative cognitive radio networks.}\label{fig1}
\end{figure}
Fig. \ref{fig1} depicts the considered system, in which a far located legitimate monitor M attempts to eavesdrop on a suspicious communication link from source S to destination D with the aid of an AF FD secondary transmitter T\footnote{With the co-existence of primary users, a priori identification of the suspicious user is required. In particular, this can be implemented by collecting massive data from the wireless mobile networks and then applying data analytics \cite{xu2017surveillance,yan2018big} such as text mining, multimedia data analysis, user mobility profiling, and social network map analysis. We would like to emphasize that our proposed energy efficient design for proactive eavesdropping is still applicable once the suspicious user is identified. There would be performance degradation compared to non-exist primary users due to additional inter-user interference.}. We consider a case that T can simultaneously listen to the suspicious source S and forward the eavesdropped information to the monitor M. Meanwhile, T can also share the spectrum of the suspicious users and transmit its own message to the desired secondary receiver R as a reward. It is assumed that the direct eavesdropping link between the monitor M and the suspicious source S is sufficiently weak to be ignored due to strong path loss or obstacles.
We further assume that each of the source S, destination D and secondary receiver R has a single antenna, whereas the secondary transmitter T is equipped with ${N_T}$ transmitting and ${N_R}$ receiving antennas, respectively. ${N_M}$ is the number of receiver antennas at the legitimate monitor M. ${h_{DS}} \in \mathbb{C}$, ${h_{RS}} \in \mathbb{C}$, ${{\bf{h}}_{TS}} \in {\mathbb{C}^{{N_R} \times 1}}$, ${{\bf{h}}_{DT}} \in {\mathbb{C}^{1 \times {N_T}}}$, ${{\bf{h}}_{RT}} \in {\mathbb{C}^{1 \times {N_T}}}$ and ${{\bf{H}}_{MT}} \in {\mathbb{C}^{{N_M} \times {N_T}}}$ are used to denote the S-D, S-R, S-T, T-D, T-R and T-M channels, respectively. Let ${{\bf{H}}_{TT}} \in {\mathbb{C}^{{N_R} \times {N_T}}}$ be the self-interference channel at the secondary transmitter T. All channels are assumed to undergo quasi-stationary flat fading, i.e., the channels are constant for a block of $N \gg 1$ suspicious symbol transmissions. In addition, we assume that the suspicious nodes are unaware of being eavesdropped on and only know the CSI of the suspicious link. Hence, conventional physical layer security techniques are not applied at the suspicious nodes to prevent eavesdropping.

At time instant $t$ for $t = 1, \cdots ,N$, the received signal at the secondary transmitter T is given by
\begin{align}\label{y_relay}
{{\bf{y}}_T}[t] = {{\bf{h}}_{TS}}{x_S}[t] + {{\bf{H}}_{TT}}{{\bf{x}}_T}[t] + {{\bf{n}}_T}[t],
\end{align}
where ${x_S}[t] \sim \mathcal{CN}(0,{P_S})$ is the transmitted signal from the suspicious source S; ${{\bf{n}}_T}[t] \sim \mathcal{CN}({\bf{0}},\sigma _T^2{{\bf{I}}_{{N_R}}})$ is the complex additive white Gaussian noise (AWGN) at T; ${{\bf{x}}_T}[t] \in {\mathbb{C}^{{N_T} \times 1}}$ is the transmitted signal at the secondary transmitter T, which takes the following form:
\begin{align}
{{\bf{x}}_T}[t] = {\bf{W}}{{\bf{y}}_T}[t - \tau ] + {\bf{v}}{s_T}[t],
\end{align}
where $\tau$ represents the processing delay at T. Herein, we consider both the NNPD (i.e., $\tau  \ge {T_S}$) and NPD (i.e., $\tau  <  < {T_S}$) cases at T, where ${T_S}$ is the period of the transmitted symbol from the suspicious source S. ${\bf{W}} \in {\mathbb{C}^{{N_T} \times {N_R}}}$ is the AF relay matrix employed at T, and ${\bf{v}} \in {\mathbb{C}^{{N_T} \times 1}}$ is the precoding vector at T for sending secondary signal ${s_T}[t]$ to the desired secondary receiver R, where $\mathbb{E}[ {{{\left| {{s_T}[t]} \right|}^2}} ] = 1$. Note that the second term in (1) is the self-interference (SI) induced by FD operation at the secondary transmitter T, which can be canceled by using zero-forcing (ZF) beamforming with either more transmit or receive antennas. In particular, the ZF constraints may take the following two forms
\begin{equation}
\begin{cases}
{\bf{W}}{{\bf{H}}_{TT}} = {\bf{0}}, &{N_R} > {N_T},\\
{{\bf{H}}_{TT}}{\bf{W}} = {\bf{0}}, &{N_T} > {N_R}.
\end{cases}
\end{equation}
Here, we focus on the case of ${N_T} > {N_R}$, as the other case can be handled similarly. Therefore, the transmit power at the secondary transmitter T can be shown to be
\begin{align}
{P_T}\left( {{\bf{W}},{\bf{v}}} \right) = {P_S}{\left\| {{\bf{W}}{{\bf{h}}_{TS}}} \right\|^2} + \sigma _T^2\left\| {\bf{W}} \right\|_F^2 + {\left\| {\bf{v}} \right\|^2}.
\end{align}

After the SI cancelation, the received signals at the suspicious destination D, the secondary receiver R, and the legitimate monitor M can be expressed, respectively, as
\begin{align}
{y_D}[t] =& {h_{DS}}{x_S}[t] + {{\bf{h}}_{DT}}{{\bf{x}}_T}[t] + {n_D}[t]\nonumber\\
 =& {h_{DS}}{x_S}[t] + {{\bf{h}}_{DT}}{\bf{W}}\left( {{{\bf{h}}_{TS}}{x_S}[t - \tau ] + {{\bf{n}}_T}[t - \tau ]} \right) \nonumber\\
 &+ {{\bf{h}}_{DT}}{\bf{v}}{s_T}[t] + {n_D}[t],
\end{align}
\begin{align}
{y_R}[t] =& {{\bf{h}}_{RT}}{{\bf{x}}_T}[t] + {h_{RS}}{x_S}[t] + {n_R}[t]\nonumber\\
 =& {{\bf{h}}_{RT}}{\bf{v}}{s_T}[t] +{{\bf{h}}_{RT}}{\bf{W}}\left( {{{\bf{h}}_{TS}}{x_S}[t - \tau ] + {{\bf{n}}_T}[t - \tau ]} \right) \nonumber\\
 &+ {h_{RS}}{x_S}[t] + {n_R}[t],
\end{align}
\begin{align}
{y_M}[t] =& {{\bf{u}}^H}{{\bf{H}}_{MT}}{{\bf{x}}_T}[t] + {{\bf{u}}^H}{{\bf{n}}_M}[t]\nonumber\\
 = &{{\bf{u}}^H}{{\bf{H}}_{MT}}{\bf{W}}\left( {{{\bf{h}}_{TS}}{x_S}[t - \tau ] + {{\bf{n}}_T}[t - \tau ]} \right) \nonumber\\&+ {{\bf{u}}^H}{{\bf{H}}_{MT}}{\bf{v}}{s_T}[t] + {{\bf{u}}^H}{{\bf{n}}_M}[t],
\end{align}
where ${n_D}[t] \sim \mathcal{CN}(0,\sigma _D^2)$, ${n_R}[t] \sim \mathcal{CN}(0,\sigma _R^2)$ and ${{\bf{n}}_M}[t] \sim \mathcal{CN}({\bf{0}},\sigma _M^2{{\bf{I}}_{{N_M}}})$ are complex AWGN at the suspicious destination D, the secondary receiver R, and the legitimate monitor M, respectively. Meanwhile, the monitor M applies the receiver combining vector ${\bf{u}} \in {\mathbb{C}^{{N_M} \times 1}}$ to decode the eavesdropped information.

As a result, the achievable rates (nats/s/Hz) at the suspicious destination D, the secondary receiver R, and the legitimate monitor M are, respectively, given by
\begin{equation}
{R_D}\left( {{\bf{W}},{\bf{v}}} \right) =
\begin{cases}
\ln \left( {1 + \frac{{{P_S}{{\left| {{h_{DS}}} \right|}^2}}}{{{J_D}\left( {{\bf{W}},{\bf{v}}} \right)}}} \right), &\text{for NNPD},\\
\ln \left( {1 + \frac{{{P_S}{{\left| {{{\bar h}_{DS}}\left( {\bf{W}} \right)} \right|}^2}}}{{{{\bar J}_D}\left( {{\bf{W}},{\bf{v}}} \right)}}} \right), &\text{for NPD},
\end{cases}
\end{equation}
\begin{align}
{R_R}\left( {{\bf{W}},{\bf{v}}} \right) = \ln \left( {1 + \frac{{{{\left| {{{\bf{h}}_{RT}}{\bf{v}}} \right|}^2}}}{{{J_R}\left( {\bf{W}} \right)}}} \right),
\end{align}
and
\begin{align}
{R_M}\left( {{\bf{W}},{\bf{v}},{\bf{u}}} \right) = \ln \left( {1 + \frac{{{P_S}{{\left| {{{\bf{u}}^H}{\bf{A(W)}}} \right|}^2}}}{{{J_M}\left( {{\bf{W}},{\bf{v}},{\bf{u}}} \right)}}} \right),
\end{align}
where ${J_D}\left( {{\bf{W}},{\bf{v}}} \right) = {P_S}{\left| {{{\bf{h}}_{DT}}{\bf{W}}{{\bf{h}}_{TS}}} \right|^2} + \sigma _T^2{\left\| {{{\bf{h}}_{DT}}{\bf{W}}} \right\|^2} + {\left| {{{\bf{h}}_{DT}}{\bf{v}}} \right|^2} + \sigma _D^2$, ${{\bar h}_{DS}}\left( {\bf{W}} \right) = {h_{DS}} + {{\bf{h}}_{DT}}{\bf{W}}{{\bf{h}}_{TS}}$, ${{\bar J}_D}\left( {{\bf{W}},{\bf{v}}} \right) = \sigma _T^2{\left\| {{{\bf{h}}_{DT}}{\bf{W}}} \right\|^2} + {\left| {{{\bf{h}}_{DT}}{\bf{v}}} \right|^2} + \sigma _D^2$, ${J_R}\left( {\bf{W}} \right) = {P_S}{\left| {{{\bf{h}}_{RT}}{\bf{W}}{{\bf{h}}_{TS}}} \right|^2} + \sigma _T^2{\left\| {{{\bf{h}}_{RT}}{\bf{W}}} \right\|^2} + {P_S}{\left| {{h_{RS}}} \right|^2} + \sigma _R^2$, ${\bf{A(W)}} = {{\bf{H}}_{MT}}{\bf{W}}{{\bf{h}}_{TS}}$ and ${J_M}\left( {{\bf{W}},{\bf{v}},{\bf{u}}} \right) = \sigma _T^2{\left\| {{{\bf{u}}^H}{{\bf{H}}_{MT}}{\bf{W}}} \right\|^2} + {\left| {{{\bf{u}}^H}{{\bf{H}}_{MT}}{\bf{v}}} \right|^2} + \sigma _M^2{\left\| {\bf{u}} \right\|^2}$.

In the sequel, we first propose an energy efficient design for proactive eavesdropping by assuming the available CSIs are perfect. We then extend our study for robust energy efficient design under imperfect CSIs related to the suspicious link. In general, the perfect CSI scenario can serve as a performance upper bound for the imperfect CSI scenario. Particularly, the imperfect CSIs can lead to substantial performance loss if not taken care of properly. As the formulated problems and solutions are exactly different, it is natural to consider the cases of perfect and imperfect CSIs separately through pertinent system designs.

\section{Energy Efficient Design Based on Perfect CSI}\label{Perfect_CSI}
In this section, we first formulate the NEE maximization problem based on the assumption that all available CSIs are perfect. Then, we propose an efficient iterative path-following algorithm to solve this non-convex problem. We further show that the proposed algorithm is guaranteed to converge with low complexity that only involves a simple convex quadratic program at each iteration.

\subsection{Problem Statement}
In our eavesdropping system, we must ensure that the eavesdropping channel capacity is greater than or equal to the data rate of the suspicious nodes so that the forwarded suspicious information via secondary transmitter T can be successfully decoded at legitimate monitor M with arbitrarily small error probability.

Based on the above analysis and perfect CSI assumption, we can express the effective eavesdropping rate as ${R_e}\left( {{\bf{W}},{\bf{v}},{\bf{u}}} \right) = {R_D}\left( {{\bf{W}},{\bf{v}}} \right)$ if ${R_M}\left( {{\bf{W}},{\bf{v}},{\bf{u}}} \right) \ge {R_D}\left( {{\bf{W}},{\bf{v}}} \right)$, and ${R_e}\left( {{\bf{W}},{\bf{v}},{\bf{u}}} \right) = 0$ otherwise. Our aim is to maximize the NEE via jointly optimizing the AF relay matrix ${\bf{W}}$ and the precoding vector ${\bf{v}}$ at the secondary transmitter T, as well as the receiver combining vector ${\bf{u}}$ at the legitimate monitor M subject to the maximum power constraint at the secondary transmitter T and the minimum data rate requirement of the SU. We also guarantee that the achievable data rate at the legitimate monitor M should be no less than the achievable data rate at suspicious destination D for efficient eavesdropping. As a result, by accounting for all of the above-mentioned factors, an NEE maximization problem is formulated as follows
\begin{subequations}\label{original_1}
\begin{align}\mathop {\max }\limits_{{\bf{W}},{\bf{v}},{\bf{u}}}\quad &\mathop \eta ( {{\bf{W}},{\bf{v}}} ) = \frac{{{\alpha _D}{R_D}\left( {{\bf{W}},{\bf{v}}} \right) + {\alpha _R}{R_R}\left( {{\bf{W}},{\bf{v}}} \right)}}{{Q( {{\bf{W}},{\bf{v}}} )}}\label{P1_obj}\\
\text{s.t.}\quad &{R_M}\left( {{\bf{W}},{\bf{v}},{\bf{u}}} \right) \ge {R_D}\left( {{\bf{W}},{\bf{v}}} \right), \label{P1_C1}\\
 &{R_R}\left( {{\bf{W}},{\bf{v}}} \right) \ge {R_{th}},\label{P1_C2}\\
&{P_S}{\left\| {{\bf{W}}{{\bf{h}}_{TS}}} \right\|^2} + \sigma _T^2\left\| {\bf{W}} \right\|_F^2 + {\left\| {\bf{v}} \right\|^2} \le {P_{\max }},\label{P1_C3}\\
&{{\bf{H}}_{TT}}{\bf{W}} = {\bf{0}},\label{P1_C4}\\
&{\left\| {\bf{u}} \right\|^2} = 1,\label{P1_C5}
\end{align}
\end{subequations}
where the constant weight factors ${\alpha _D} \ge 0$ and ${\alpha _R} \ge 0$ reflect the priorities of the eavesdropping rate and the SU's achievable rate. $Q\left( {{\bf{W}},{\bf{v}}} \right) = {{{P_T}\left( {{\bf{W}},{\bf{v}}} \right)} \mathord{\left/
 {\vphantom {{{P_T}\left( {{\bf{W}},{\bf{v}}} \right)} \xi }} \right.
 \kern-\nulldelimiterspace} \xi } + {N_T}{P_A} + {N_R}{P_R} + {P_C}$ is the overall energy consumption of the secondary transmitter T. $\xi  \in \left( {0,1} \right]$, ${P_A}$, ${P_R}$, and ${P_C}$ are the power amplifier efficiency, power dissipation at each transmitting and receiving antenna, and constant circuit power consumption of T, respectively. (\ref{P1_C1}) guarantees that the suspicious link can be successfully eavesdropped by monitor M. (\ref{P1_C2}) ensures that the minimum data rate requirement ${R_{th}}$ of SU can be satisfied; otherwise, the secondary transmitter T may not to help the monitor. ${P_{\max }}$ represents the maximum available power at the secondary transmitter T, and (\ref{P1_C4}) guarantees the ZF constraint to cancel SI at the secondary transmitter T.
\begin{remark}
The NEE can be further denoted as the weighted sum of the effective eavesdropping EE and SU's EE, i.e., $\eta \left( {{\bf{W}},{\bf{v}}} \right) = {\alpha _D}{\eta _D}\left( {{\bf{W}},{\bf{v}}} \right) + {\alpha _R}{\eta _R}\left( {{\bf{W}},{\bf{v}}} \right)$, where ${\eta _D}\left( {{\bf{W}},{\bf{v}}} \right) = {{{R_D}\left( {{\bf{W}},{\bf{v}}} \right)} \mathord{\left/
 {\vphantom {{{R_D}\left( {{\bf{W}},{\bf{v}}} \right)} {Q\left( {{\bf{W}},{\bf{v}}} \right)}}} \right.
 \kern-\nulldelimiterspace} {Q\left( {{\bf{W}},{\bf{v}}} \right)}}$ and ${\eta _R}\left( {{\bf{W}},{\bf{v}}} \right) = {{{R_R}\left( {{\bf{W}},{\bf{v}}} \right)} \mathord{\left/
 {\vphantom {{{R_R}\left( {{\bf{W}},{\bf{v}}} \right)} {Q\left( {{\bf{W}},{\bf{v}}} \right)}}} \right.
 \kern-\nulldelimiterspace} {Q\left( {{\bf{W}},{\bf{v}}} \right)}}$. Therefore, the NEE and weighted sum energy efficiency (WSEE) are equivalent in our scenario.
\end{remark}
We observe that only ${R_M}\left( {{\bf{W}},{\bf{v}},{\bf{u}}} \right)$ in (\ref{P1_C1}) depends on ${\bf{u}}$. Accordingly, for a given ${\bf{W}}$ and ${\bf{v}}$, the optimal ${\bf{u}}$ is obtained by maximizing the received signal-to-interference-plus-noise ratio (SINR) at the monitor M, i.e., $\frac{{{P_S}{{\left| {{{\bf{u}}^H}{\bf{A(W)}}} \right|}^2}}}{{{J_M}\left( {{\bf{W}},{\bf{v}},{\bf{u}}} \right)}}$, which can be rewritten as a generalized Rayleigh quotient problem
\begin{align}
\mathop {\max }\limits_{{{\left\| {\bf{u}} \right\|}^2} = 1} \frac{{{P_S}{{\bf{u}}^H}{\bf{A(W)A(W}}{{\bf{)}}^H}{\bf{u}}}}{{{{\bf{u}}^H}{\bf{\Phi }}\left( {{\bf{W}},{\bf{v}}} \right){\bf{u}}}},
\end{align}
where ${\bf{\Phi }}\left( {{\bf{W}},{\bf{v}}} \right) = \sigma _T^2{{\bf{H}}_{MT}}{\bf{W}}{{\bf{W}}^H}{\bf{H}}_{MT}^H + {{\bf{H}}_{MT}}{\bf{v}}{{\bf{v}}^H}{\bf{H}}_{MT}^H + \sigma _M^2{\bf{I}}$. Therefore, the optimal ${\bf{u}}$ can be derived as
\begin{align}
{{\bf{u}}^*} = \frac{{{\bf{\Phi }}{{\left( {{\bf{W}},{\bf{v}}} \right)}^{ - 1}}{\bf{A(W)}}}}{{\left\| {{\bf{\Phi }}{{\left( {{\bf{W}},{\bf{v}}} \right)}^{ - 1}}{\bf{A(W)}}} \right\|}}.
\end{align}
Now, we can rewrite ${R_M}\left( {{\bf{W}},{\bf{v}},{\bf{u}}} \right)$ below by substituting ${{\bf{u}}^*}$,
\begin{align}
{R_M}\left( {{\bf{W}},{\bf{v}}} \right) = \ln (1 + {P_S}{\bf{A(W}}{{\bf{)}}^H}{\bf{\Phi }}{\left( {{\bf{W}},{\bf{v}}} \right)^{ - 1}}{\bf{A(W)}}).
\end{align}

In addition, we further consider the ZF constraint in (\ref{P1_C4}). For simplicity, we define ${\bf{W}} = {{\bf{V}}_{\bf{0}}}{\bf{G}}$, where ${{\bf{V}}_{\bf{0}}} \in {\mathbb{C}^{{N_T} \times ({N_T} - {N_R})}}$ is a semi-unitary matrix lies in the null space of ${{\bf{H}}_{TT}}$, which can be the right singular vectors associated with the zero singular values of ${{\bf{H}}_{TT}}$. ${\bf{G}} \in {\mathbb{C}^{({N_T} - {N_R}) \times {N_R}}}$ is a new optimization matrix to be designed. Therefore, problem (\ref{original_1}) reduces to
\begin{subequations}\label{pro_1}
\begin{align}\mathop {\max }\limits_{{\bf{G}},{\bf{v}}}\quad &\mathop \eta \left( {{\bf{G}},{\bf{v}}} \right)  \buildrel \Delta \over =  {\alpha _D}{\eta _D}\left( {{\bf{G}},{\bf{v}}} \right) + {\alpha _R}{\eta _R}\left( {{\bf{G}},{\bf{v}}} \right)\label{P2_obj}\\
\text{s.t.}\quad &{R_M}\left( {{\bf{G}},{\bf{v}}} \right) \ge {R_D}\left( {{\bf{G}},{\bf{v}}} \right), \label{P2_C1}\\
 &{R_R}\left( {{\bf{G}},{\bf{v}}} \right) \ge {R_{th}},\label{P2_C2}\\
&{P_S}{\left\| {{\bf{G}}{{\bf{h}}_{TS}}} \right\|^2} + \sigma _T^2\left\| {\bf{G}} \right\|_F^2 + {\left\| {\bf{v}} \right\|^2} \le {P_{\max }}.\label{P2_C3}
\end{align}
\end{subequations}

We note that problem (\ref{pro_1}) is a non-convex problem due to the non-concave objective function, as well as the non-convexity of the constraints (\ref{P2_C1}) and (\ref{P2_C2}), which is quite complex and hard to solve even to find a feasible solution. There is no benefit via using conventional Dinkelbach's algorithm \cite{dinkelbach1967nonlinear,cheung2013achieving} since the problem is still as non-convex and computationally difficult as its original problem (\ref{pro_1}). Instead, we attempt to develop an efficient iterative path-following algorithm to solve the problem, which will be explained more clearly in the next subsection. Note that our work is motivated by the fact that the path-following algorithm has already demonstrated superior performance over other extant algorithms in the literature \cite{nasir2017secure,nguyen2018new}.

\subsection{Proposed Iterative Path-Following Algorithm}\label{path_following}
In this part, we first find a lower bounding concave approximation for objective function (\ref{P2_obj}) and the inner convex approximations for constraints (\ref{P2_C1}) and (\ref{P2_C2}). Then, we propose an efficient iterative path-following algorithm for solving the non-convex problem. In addition, we develop an efficient scheme to identify a feasible initial solution to execute the algorithm.

In order to develop our path-following iterations for the solution of problem (\ref{pro_1}), we introduce the following lemmas.
\begin{lemma}\label{L_1}
For every $x > 0$ and given $\bar x > 0$, we have $\ln (1 + x) \le \ln (1 + \bar x) + {{(x - \bar x)} \mathord{\left/
 {\vphantom {{(x - \bar x)} {(1 + \bar x)}}} \right.
 \kern-\nulldelimiterspace} {(1 + \bar x)}}$.
\end{lemma}
\begin{lemma}\label{L_2}
For every $x > 0,y > 0,$ and given $\bar x > 0$ and $\bar y > 0$, we have
$\ln \left( {1 + \frac{1}{{xy}}} \right) \ge \ln \left( {1 + \frac{1}{{\bar x\bar y}}} \right) + \frac{{1/\bar x\bar y}}{{1 + 1/\bar x\bar y}}\left( {2 - \frac{x}{{\bar x}} - \frac{y}{{\bar y}}} \right)$.
\end{lemma}
\begin{lemma}\label{L_3}
The following inequality holds: $\frac{{\ln \left( {1 + x} \right)}}{y} \ge 2\frac{{\ln \left( {1 + \bar x} \right)}}{{\bar y}} + \frac{{\bar x}}{{\bar y\left( {1 + \bar x} \right)}} - \frac{{{{\bar x}^2}}}{{\bar y\left( {1 + \bar x} \right)}}\frac{1}{x} - \frac{{\ln \left( {1 + \bar x} \right)}}{{{{\bar y}^2}}}y$ for all $x > 0,y > 0,\bar x > 0$ and $\bar y > 0$.
\end{lemma}
\begin{lemma}\label{L_4}
${\left\| {\bf{x}} \right\|^2} \ge 2{\mathop{\Re}\nolimits} \left\{ {{{{\bf{\bar x}}}^H}{\bf{x}}} \right\} - {\left\| {{\bf{\bar x}}} \right\|^2},\forall {\bf{x}} \in {\mathbb{C}^m},{\bf{\bar x}} \in {\mathbb{C}^m}$.
\end{lemma}

Given that for a convex function $f(x)$, the Taylor approximation at any point is always its global under-estimator, i.e., $f(x) \ge f(\bar x) + \nabla f{(\bar x)^H}(x - \bar x)$, where $\nabla f( \bullet )$ is the gradient of $f( \bullet )$. It is easy to prove the above lemmas based on the convexity of functions $- \ln (1 + x)$, $\ln \left( {1 + \frac{1}{{xy}}} \right)$ and $\frac{{\ln \left( {1 + {1 \mathord{\left/
 {\vphantom {1 x}} \right.
 \kern-\nulldelimiterspace} x}} \right)}}{y}$ in the domain $x > 0,y > 0$, as well as the convex function ${\left\| {\bf{x}} \right\|^2}$ for any ${\bf{x}} \in {\mathbb{C}^m}$.
We now give the main steps to approximately transform the non-convex functions of problem (15) into the convex forms by using these lemmas. To avoid overwhelming chaotic, we only present the key points and final results as below.

Employing \textbf{Lemma \ref{L_3}} and \textbf{\ref{L_4}}, we have
\begin{align}
{\eta _D}\left( {{\bf{G}},{\bf{v}}} \right) \ge {g^{(\ell)}}({\bf{G}},{\bf{v}}),
\end{align}
\begin{align}
{\eta _R}\left( {{\bf{G}},{\bf{v}}} \right) \ge {\pi ^{(\ell)}}({\bf{G}},{\bf{v}}),
\end{align}
where
\begin{equation*}
{g^{(\ell)}}({\bf{G}},{\bf{v}}) \buildrel \Delta \over =
\begin{cases}
{\lambda ^{(\ell)}} - \frac{{{\sigma ^{(\ell)}}{J_D}\left( {{\bf{G}},{\bf{v}}} \right)}}{{{P_S}{{\left| {{h_{DS}}} \right|}^2}}} - {\mu ^{(\ell)}}Q\left( {{\bf{G}},{\bf{v}}} \right), \text{for NNPD},\\
{{\bar \lambda }^{(\ell)}} - \frac{{{{\bar \sigma }^{(\ell)}}{{\bar J}_D}\left( {{\bf{G}},{\bf{v}}} \right)}}{{{P_S}{{\bar \delta }_{DS}}({\bf{G}})}} - {{\bar \mu }^{(\ell)}}Q\left( {{\bf{G}},{\bf{v}}} \right), \text{for NPD},
\end{cases}
\end{equation*}
${\lambda ^{(\ell)}} = 2\frac{{\ln (1 + \gamma _D^{(\ell)})}}{{Q({{\bf{G}}^{(\ell)}},{{\bf{v}}^{(\ell)}})}} + \frac{{\gamma _D^{(\ell)}}}{{Q({{\bf{G}}^{(\ell)}},{{\bf{v}}^{(\ell)}})(1 + \gamma _D^{(\ell)})}},{\sigma ^{(\ell)}} = \frac{{{{(\gamma _D^{(\ell)})}^2}}}{{Q({{\bf{G}}^{(\ell)}},{{\bf{v}}^{(\ell)}})(1 + \gamma _D^{(\ell)})}},{\mu ^{(\ell)}} = \frac{{\ln (1 + \gamma _D^{(\ell)})}}{{Q{{({{\bf{G}}^{(\ell)}},{{\bf{v}}^{(\ell)}})}^2}}},\gamma _D^{(\ell)} = \frac{{{P_S}{{\left| {{h_{DS}}} \right|}^2}}}{{{J_D}({{\bf{G}}^{(\ell)}},{{\bf{v}}^{(\ell)}})}},$
${{\bar \lambda }^{(\ell)}} = 2\frac{{\ln (1 + \bar \gamma _D^{(\ell)})}}{{Q({{\bf{G}}^{(\ell)}},{{\bf{v}}^{(\ell)}})}} + \frac{{\bar \gamma _D^{(\ell)}}}{{Q({{\bf{G}}^{(\ell)}},{{\bf{v}}^{(\ell)}})(1 + \bar \gamma _D^{(\ell)})}},{{\bar \sigma }^{(\ell)}} = \frac{{{{(\bar \gamma _D^{(\ell)})}^2}}}{{Q({{\bf{G}}^{(\ell)}},{{\bf{v}}^{(\ell)}})(1 + \bar \gamma _D^{(\ell)})}},{{\bar \mu }^{(\ell)}} = \frac{{\ln (1 + \bar \gamma _D^{(\ell)})}}{{Q{{({{\bf{G}}^{(\ell)}},{{\bf{v}}^{(\ell)}})}^2}}},\bar \gamma _D^{(\ell)} = \frac{{{P_S}{{\left| {{{\bar h}_{DS}}\left( {{{\bf{G}}^{(\ell)}}} \right)} \right|}^2}}}{{{{\bar J}_D}({{\bf{G}}^{(\ell)}},{{\bf{v}}^{(\ell)}})}},$ and
${{\bar \delta }_{DS}}({\bf{G}}) = 2\mathop{\Re}\left\{ {{{\left( {{{\bar h}_{DS}}\left( {{{\bf{G}}^{(\ell)}}} \right)} \right)}^H}\left( {{{\bar h}_{DS}}\left( {\bf{G}} \right)} \right)} \right\} - {\left\| {{{\bar h}_{DS}}\left( {{{\bf{G}}^{(\ell)}}} \right)} \right\|^2} \ge 0$.

${\pi ^{(\ell)}}({\bf{G}},{\bf{v}}) \buildrel \Delta \over = {\theta ^{(\ell)}} - {\varepsilon ^{(\ell)}}\frac{{{J_R}\left( {\bf{G}} \right)}}{{{\delta _R}({\bf{v}})}} - {\varphi ^{(\ell)}}Q\left( {{\bf{G}},{\bf{v}}} \right),$ ${\theta ^{(\ell)}} = 2\frac{{\ln (1 + \gamma _R^{(\ell)})}}{{Q({{\bf{G}}^{(\ell)}},{{\bf{v}}^{(\ell)}})}} + \frac{{\gamma _R^{(\ell)}}}{{Q({{\bf{G}}^{(\ell)}},{{\bf{v}}^{(\ell)}})(1 + \gamma _R^{(\ell)})}},{\varepsilon ^{(\ell)}} = \frac{{{{(\gamma _R^{(\ell)})}^2}}}{{Q({{\bf{G}}^{(\ell)}},{{\bf{v}}^{(\ell)}})(1 + \gamma _R^{(\ell)})}},{\varphi ^{(\ell)}} = \frac{{\ln (1 + \gamma _R^{(\ell)})}}{{Q{{({{\bf{G}}^{(\ell)}},{{\bf{v}}^{(\ell)}})}^2}}},\gamma _R^{(\ell)} = \frac{{{{\left| {{{\bf{h}}_{RT}}{{\bf{v}}^{(\ell)}}} \right|}^2}}}{{{J_R}\left( {{{\bf{G}}^{(\ell)}}} \right)}}$, and ${\delta _R}({\bf{v}}) = 2{\mathop{\Re}\nolimits}\{ {({{\bf{v}}^{(\ell)}})^H}{\bf{h}}_{RT}^H{{\bf{h}}_{RT}}{\bf{v}}\}  - {\left\| {{{\bf{h}}_{RT}}{{\bf{v}}^{(\ell)}}} \right\|^2} \ge 0$.

Combining \textbf{Lemma \ref{L_2}} and \textbf{\ref{L_4}}, we have
\begin{align}
{R_R}\left( {{\bf{G}},{\bf{v}}} \right) \ge {\upsilon ^{(\ell)}}({\bf{G}},{\bf{v}}),
\end{align}
where ${\upsilon ^{(\ell)}}({\bf{G}},{\bf{v}}) \buildrel \Delta \over = \ln (1 + \gamma _R^{(\ell)}) + \frac{{\gamma _R^{(\ell)}}}{{1 + \gamma _R^{(\ell)}}}\left( {2 - \frac{{{J_R}\left( {\bf{G}} \right)}}{{{J_R}\left( {{{\bf{G}}^{(\ell)}}} \right)}} - \frac{{{{\left| {{{\bf{h}}_{RT}}{{\bf{v}}^{(\ell)}}} \right|}^2}}}{{{\delta _R}({\bf{v}})}}} \right)$.

Next, we introduce the following lemma to find the inner convex approximation for the left-hand side of constraint (\ref{P2_C1}).
\begin{lemma}\label{L_5}
(\cite{nguyen2018new}): For every ${\bf{x}} \in {\mathbb{C}^m}$ and positive definite matrix variable ${\bf{Y}} \in {\mathbb{C}^{m \times m}}$, given ${\bf{\bar x}} \in {\mathbb{C}^m}$ and ${\bf{\bar Y}} \in {\mathbb{C}^{m \times m}}$, we have $\ln (1 + {{\bf{x}}^H}{{\bf{Y}}^{ - 1}}{\bf{x}}) \ge \ln (1 + {{{\bf{\bar x}}}^H}{{{\bf{\bar Y}}}^{ - 1}}{\bf{\bar x}}) - {{{\bf{\bar x}}}^H}{{{\bf{\bar Y}}}^{ - 1}}{\bf{\bar x}} + 2{\mathop{\Re}\nolimits}\{ {{{\bf{\bar x}}}^H}{{{\bf{\bar Y}}}^{ - 1}}{\bf{x}}\}  - \text{tr}\{ {[{{{\bf{\bar Y}}}^{ - 1}} - {({\bf{\bar Y}} + {\bf{\bar x}}{{{\bf{\bar x}}}^H})^{ - 1}}]^H}({\bf{x}}{{\bf{x}}^H} + {\bf{Y}})\}$.
\end{lemma}

Based on \textbf{Lemma \ref{L_5}}, we can obtain
\begin{align}
{R_M}\left( {{\bf{G}},{\bf{v}}} \right) \ge {\rho ^{(\ell)}}({\bf{G}},{\bf{v}}),
\end{align}
where ${\rho ^{(\ell)}}({\bf{G}},{\bf{v}}) \buildrel \Delta \over = \ln (1 + \gamma _M^{(\ell)}) - \gamma _M^{(\ell)} + 2{\mathop{\Re}\nolimits}\{ {P_S}{\bf{A(}}{{\bf{G}}^{(\ell)}}{{\bf{)}}^H}{({\bf{\Phi }}({{\bf{G}}^{(\ell)}},{{\bf{v}}^{(\ell)}}))^{ - 1}}{\bf{A(G)}}\}  - \text{tr}\{ [{({\bf{\Phi }}({{\bf{G}}^{(\ell)}},{{\bf{v}}^{(\ell)}}))^{ - 1}} - ({\bf{\Phi }}({{\bf{G}}^{(\ell)}},{{\bf{v}}^{(\ell)}}) + $ ${P_S}{\bf{A(}}{{\bf{G}}^{(\ell)}}{\bf{)A(}}{{\bf{G}}^{(\ell)}}{{\bf{)}}^H}{)^{ - 1}}{]^H}({P_S}{\bf{A(G)A(G}}{{\bf{)}}^H} + {\bf{\Phi }}({\bf{G}},{\bf{v}}))\} $, and $\gamma _M^{(\ell)} = {P_S}{\bf{A(}}{{\bf{G}}^{(\ell)}}{{\bf{)}}^H}{({\bf{\Phi }}({{\bf{G}}^{(\ell)}},{{\bf{v}}^{(\ell)}}))^{ - 1}}{\bf{A(}}{{\bf{G}}^{(\ell)}}{\bf{)}}$.

For the right-hand side of constraint (\ref{P2_C1}), we apply \textbf{Lemma \ref{L_1}} and \textbf{\ref{L_4}}. Consequently, the following inequality holds:
\begin{align}
{R_D}\left( {{\bf{G}},{\bf{v}}} \right) \le {\varsigma ^{(\ell)}}({\bf{G}},{\bf{v}}),
\end{align}
where
\begin{equation*}
{\varsigma ^{(\ell)}}({\bf{G}},{\bf{v}}) \buildrel \Delta \over =
\begin{cases}
\ln (1 + \gamma _D^{(\ell)}) + \frac{{\chi ({\bf{G}},{\bf{v}}) - \gamma _D^{(\ell)}}}{{1 + \gamma _D^{(\ell)}}}, &\text{for NNPD},\\
\ln (1 + \bar \gamma _D^{(\ell)}) + \frac{{\bar \chi ({\bf{G}},{\bf{v}}) - \bar \gamma _D^{(\ell)}}}{{1 + \bar \gamma _D^{(\ell)}}}, &\text{for NPD},
\end{cases}
\end{equation*}
$\chi ({\bf{G}},{\bf{v}}) = \frac{{{P_S}{{\left| {{h_{DS}}} \right|}^2}}}{{{P_S}{\delta _{DS}}({\bf{G}}) + \sigma _T^2{\delta _{DT}}({\bf{G}}) + {\delta _D}({\bf{v}}) + \sigma _D^2}}$, $\bar \chi ({\bf{G}},{\bf{v}}) = \frac{{{P_S}{{\left| {{{\bar h}_{DS}}\left( {\bf{G}} \right)} \right|}^2}}}{{\sigma _T^2{\delta _{DT}}({\bf{G}}) + {\delta _D}({\bf{v}}) + \sigma _D^2}}$,
${\delta _{DS}}({\bf{G}}) = 2{\mathop{\Re}\nolimits}\{ {\bf{h}}_{TS}^H{({{\bf{V}}_{\bf{0}}}{{\bf{G}}^{(\ell)}})^H}{\bf{h}}_{DT}^H{{\bf{h}}_{DT}}{{\bf{V}}_{\bf{0}}}{\bf{G}}{{\bf{h}}_{TS}}\}  - {\left\| {{{\bf{h}}_{DT}}{{\bf{V}}_{\bf{0}}}{{\bf{G}}^{(\ell)}}{{\bf{h}}_{TS}}} \right\|^2} \ge 0$, ${\delta _{DT}}({\bf{G}}) = 2{\mathop{\Re}\nolimits}\{ {{\bf{h}}_{DT}}{{\bf{V}}_{\bf{0}}}{\bf{G}}{({{\bf{V}}_{\bf{0}}}{{\bf{G}}^{(\ell)}})^H}{\bf{h}}_{DT}^H\}  - {\left\| {{{\bf{h}}_{DT}}{{\bf{V}}_{\bf{0}}}{{\bf{G}}^{(\ell)}}} \right\|^2} \ge 0$, and ${\delta _D}({\bf{v}}) = 2{\mathop{\Re}\nolimits}\{ {({{\bf{v}}^{(\ell)}})^H}{\bf{h}}_{DT}^H{{\bf{h}}_{DT}}{\bf{v}}\}  - {\left\| {{{\bf{h}}_{DT}}{{\bf{v}}^{(\ell)}}} \right\|^2} \ge 0$.

In summary, we solve the following convex quadratic program to achieve minorant
maximization of the non-convex problem (\ref{pro_1}) at the $\ell$-th iteration:
\begin{subequations}\label{pro_2}
\begin{align}\mathop {\max }\limits_{{\bf{G}},{\bf{v}}}\quad &{\eta ^{(\ell)}}\left( {{\bf{G}},{\bf{v}}} \right) \buildrel \Delta \over = {\alpha _D}{g^{(\ell)}}({\bf{G}},{\bf{v}}) + {\alpha _R}{\pi ^{(\ell)}}({\bf{G}},{\bf{v}})\label{P3_obj}\\
\text{s.t.}\quad &{\rho ^{(\ell)}}({\bf{G}},{\bf{v}}) \ge {\varsigma ^{(\ell)}}({\bf{G}},{\bf{v}}), \label{P3_C1}\\
 &{\upsilon ^{(\ell)}}({\bf{G}},{\bf{v}}) \ge {R_{th}},\label{P3_C2}\\
&(\ref{P2_C3}).\label{P3_C3}
\end{align}
\end{subequations}

Therefore, the proposed iterative path-following algorithm used for solving problem (\ref{pro_1}) is summarized in \textbf{Algorithm \ref{alg:A}}.
\begin{algorithm}[H]
\caption{Iterative Path-Following Algorithm to Solve (\ref{pro_1})}
\label{alg:A}
\begin{algorithmic}
\STATE {Initialize $\ell: = 0$.}
\STATE {Solve (\ref{pro_3}) to generate a feasible initial point ${({{\bf{G}}^{(0)}},{{\bf{v}}^{(0)}})}$.}
\REPEAT
\STATE \begin{enumerate}
         \item Solve problem (\ref{pro_2}) to obtain ${({{\bf{G}}^{(\ell+1)}},{{\bf{v}}^{(\ell+1)}})}$;
         \item $\ell: = \ell + 1$;
       \end{enumerate}
\UNTIL{convergence of the objective in problem (\ref{pro_1}).}
\end{algorithmic}
\end{algorithm}

\textsl{Initialization of \textbf{Algorithm \ref{alg:A}}:} To find a feasible initial point for (\ref{pro_1}), we iteratively solve the following convex problem:
\begin{align}\label{pro_3}
\mathop {\max }\limits_{{\bf{G}},{\bf{v}}}\quad t, \quad\quad\quad \text{s.t.}\quad (\ref{P2_C3}),
\end{align}
where $t = \min \{ {\rho ^{(\ell)}}({\bf{G}},{\bf{v}}) - {\varsigma ^{(\ell)}}({\bf{G}},{\bf{v}}),{\upsilon ^{(\ell)}}({\bf{G}},{\bf{v}}) - {R_{th}}\}$. A feasible initial point to execute \textbf{Algorithm \ref{alg:A}} can be the solution of problem (\ref{pro_3}) whenever $t \ge 0$.


\subsection{Convergence and Complexity Analysis}

\textsl{Convergence Analysis:} Since problem (\ref{pro_2}) is a minorant maximization of problem (\ref{pro_1}), we have
$\eta \left( {{\bf{G}},{\bf{v}}} \right) \ge {\eta ^{(\ell)}}\left( {{\bf{G}},{\bf{v}}} \right),\forall {\bf{G}},{\bf{v}}$
and $\eta \left( {{{\bf{G}}^{(\ell)}},{{\bf{v}}^{(\ell)}}} \right) = {\eta ^{(\ell)}}\left( {{{\bf{G}}^{(\ell)}},{{\bf{v}}^{(\ell)}}} \right)$. Therefore, $\eta \left( {{{\bf{G}}^{(\ell + 1)}},{{\bf{v}}^{(\ell + 1)}}} \right) \ge {\eta ^{(\ell)}}\left( {{{\bf{G}}^{(\ell + 1)}},{{\bf{v}}^{(\ell + 1)}}} \right) > {\eta ^{(\ell)}}\left( {{{\bf{G}}^{(\ell)}},{{\bf{v}}^{(\ell)}}} \right) = \eta \left( {{{\bf{G}}^{(\ell)}},{{\bf{v}}^{(\ell)}}} \right)$ as far as $\{ {{\bf{G}}^{(\ell + 1)}},{{\bf{v}}^{(\ell + 1)}}\}  \ne \{ {{\bf{G}}^{(\ell)}},{{\bf{v}}^{(\ell)}}\}$, where the second inequality follows from the fact that $\{ {{\bf{G}}^{(\ell + 1)}},{{\bf{v}}^{(\ell + 1)}}\}$ and $\{ {{\bf{G}}^{(\ell)}},{{\bf{v}}^{(\ell)}}\}$ are the optimal and a feasible point of problem (\ref{pro_2}), respectively. This result shows that $\{ {{\bf{G}}^{(\ell + 1)}},{{\bf{v}}^{(\ell + 1)}}\}$ is better than $\{ {{\bf{G}}^{(\ell)}},{{\bf{v}}^{(\ell)}}\}$ for problem (\ref{pro_1}) whenever $\{ {{\bf{G}}^{(\ell + 1)}},{{\bf{v}}^{(\ell + 1)}}\}  \ne \{ {{\bf{G}}^{(\ell)}},{{\bf{v}}^{(\ell)}}\}$. On the other hand, if $\{ {{\bf{G}}^{(\ell + 1)}},{{\bf{v}}^{(\ell + 1)}}\}  = \{ {{\bf{G}}^{(\ell)}},{{\bf{v}}^{(\ell)}}\}$, i.e., $\{ {{\bf{G}}^{(\ell)}},{{\bf{v}}^{(\ell)}}\}$ is also the optimal solution of the convex problem (\ref{pro_2}), then it must satisfy the first-order necessary optimality condition of problem (\ref{pro_2}), which obviously is also the first-order necessary optimality condition of problem (\ref{pro_1}). Therefore, we conclude that the proposed path-following algorithm generates a non-decreasing sequence of objective values for problem (\ref{pro_1}), i.e., $\eta \left( {{{\bf{G}}^{(\ell + 1)}},{{\bf{v}}^{(\ell + 1)}}} \right) \ge \eta \left( {{{\bf{G}}^{(\ell)}},{{\bf{v}}^{(\ell)}}} \right)$. Furthermore, the sequence $\{ ({{\bf{G}}^{(\ell)}},{{\bf{v}}^{(\ell)}})\}$ converges at least to a locally optimal solution satisfying the first-order necessary optimality condition of the non-convex problem (\ref{pro_1}).

\textsl{Complexity Analysis:} In particular, as problem (\ref{pro_2}) has $\bar n = ({N_T} - {N_R}){N_R} + {N_T}$ scalar decision variables and seven quadratic and linear constraints, its computational complexity per iteration is ${\rm \mathcal{O}}({7^{2.5}}({{\bar n}^2} + 7))$ for both the NNPD and NPD, respectively.

\section{Robust Energy Efficient Design Based on Imperfect CSI}\label{IMPerfect_CSI}
In the previous section, we assumed that the available CSIs are perfect. 
As the suspicious nodes are assumed to be unaware of being eavesdropped on, the CSIs related to the suspicious pairs can only be obtained by overhearing either pilot signals or feedbacked channel information from the suspicious nodes. Hence, the CSIs related to the suspicious pairs cannot be perfectly known since there is no cooperation among the suspicious nodes and eavesdropping nodes. However, we assume all of the other CSIs are perfect by taking advantage of sufficient cooperation and advanced channel estimation method. Note that our proposed robust NEE deign in this Section can be generalized to the scenario that all the available CSIs are imperfect in a straightforward manner.

To address this, we formulate and investigate a worst-case NEE maximization problem in this section according to the
imperfect CSIs.
Based on the adopted assumption, we capture the imperfect information of CSIs by adopting the following norm-bounded channel uncertainty model,
\begin{subequations}
\begin{align}
{h_{DS}} = {{\hat h}_{DS}} + \Delta {h_{DS}},\ \left\| {\Delta {h_{DS}}} \right\| \le {\epsilon _{DS}},\\
{{\bf{h}}_{TS}} = {{{\bf{\hat h}}}_{TS}} + {\bf{\Delta }}{{\bf{h}}_{TS}},\ \left\| {{\bf{\Delta }}{{\bf{h}}_{TS}}} \right\| \le {\epsilon _{TS}},\label{ICSI_TS}\\
{{\bf{h}}_{DT}} = {{{\bf{\hat h}}}_{DT}} + {\bf{\Delta }}{{\bf{h}}_{DT}},\ \left\| {{\bf{\Delta }}{{\bf{h}}_{DT}}} \right\| \le {\epsilon _{DT}},\\
{h_{RS}} = {{\hat h}_{RS}} + \Delta {h_{RS}},\ \left\| {\Delta {h_{RS}}} \right\| \le {\epsilon _{RS}},\label{ICSI_RS}
\end{align}
\end{subequations}
where ${{\hat h}_{DS}}$, ${{{\bf{\hat h}}}_{TS}}$, ${{{\bf{\hat h}}}_{DT}}$, and ${{\hat h}_{RS}}$ are the estimated versions of ${h_{DS}}$, ${{\bf{h}}_{TS}}$, ${{\bf{h}}_{DT}}$, and ${h_{RS}}$, respectively. $\Delta {h_{DS}}$, ${\bf{\Delta }}{{\bf{h}}_{TS}}$, ${\bf{\Delta }}{{\bf{h}}_{DT}}$, and $\Delta {h_{RS}}$ represent the corresponding channel estimation errors, which are norm bounded by the given radius ${\epsilon _{DS}}$, ${\epsilon _{TS}}$, ${\epsilon _{DT}}$, and ${\epsilon _{RS}}$, respectively. Note that the robust design with probabilistic channel model \cite{zhou2017robust,zhou2020robust} for delay-sensitive communications requires a separate study in the future.

\subsection{Problem Statement}
According to the channel uncertainties, the worst-case NEE maximization problem can be formulated as follows
\begin{subequations}\label{Robust_Original}
\begin{align}&\mathop {\max }\limits_{{\bf{W}},{\bf{v}},{\bf{u}}}\quad \mathop {\min }\limits_{\scriptstyle\Delta {h_{DS}},{\bf{\Delta }}{{\bf{h}}_{DT}}\hfill\atop
\scriptstyle{\bf{\Delta }}{{\bf{h}}_{TS}},\Delta {h_{RS}}\hfill} \frac{{{\alpha _D}{R_D}\left( {{\bf{W}},{\bf{v}}} \right) + {\alpha _R}{R_R}\left( {{\bf{W}},{\bf{v}}} \right)}}{{Q({\bf{W}},{\bf{v}})}}\\
&\text{s.t.}\quad \mathop {\min }\limits_{{\bf{\Delta }}{{\bf{h}}_{TS}}} {R_M}\left( {{\bf{W}},{\bf{v}},{\bf{u}}} \right) \ge \mathop {\max }\limits_{\Delta {h_{DS}},{\bf{\Delta }}{{\bf{h}}_{DT}},{\bf{\Delta }}{{\bf{h}}_{TS}}} {R_D}\left( {{\bf{W}},{\bf{v}}} \right), \\
 &\mathop {\min }\limits_{{\bf{\Delta }}{{\bf{h}}_{TS}},\Delta {h_{RS}}} {R_R}\left( {{\bf{W}},{\bf{v}}} \right) \ge {R_{th}},\\
&\mathop {\max }\limits_{{\bf{\Delta }}{{\bf{h}}_{TS}}} {P_S}{\left\| {{\bf{W}}{{\bf{h}}_{TS}}} \right\|^2} + \sigma _T^2\left\| {\bf{W}} \right\|_F^2 + {\left\| {\bf{v}} \right\|^2} \le {P_{\max }},\\
&{{\bf{H}}_{TT}}{\bf{W}} = {\bf{0}},\label{Robust_zf}\\
&{\left\| {\bf{u}} \right\|^2} = 1,\label{Robust_u}\\
&\left\| {\Delta {h_{DS}}} \right\| \le {\epsilon _{DS}},\left\| {{\bf{\Delta }}{{\bf{h}}_{DT}}} \right\| \le {\epsilon _{DT}},\label{ICSI_1}\\
&\left\| {{\bf{\Delta }}{{\bf{h}}_{TS}}} \right\| \le {\epsilon _{TS}},\left\| {\Delta {h_{RS}}} \right\| \le {\epsilon _{RS}}.\label{ICSI_2}
\end{align}
\end{subequations}

Naturally, the optimal solution to problem (\ref{Robust_Original}) is robust in the
presence of CSI perturbation. For simplicity, we address the ZF constraint in (\ref{Robust_zf}) in a similar manner to that in Section \ref{Perfect_CSI}, and replace ${\bf{W}}$ with a new optimization matrix ${\bf{G}} \in {\mathbb{C}^{({N_T} - {N_R}) \times {N_R}}}$ to be designed. By introducing the slack variables ${t_D}$, ${t_R}$ and ${t_S}$, we further rewrite problem (\ref{Robust_Original}) as
\begin{subequations}\label{Robust_Simplify}
\begin{align}&\mathop {\max }\limits_{{\bf{G}},{\bf{v}},{\bf{u}},{t_D}\hfill\atop
\scriptstyle{t_R},{t_S}}\quad \mathop \frac{{{t_D} + {t_R}}}{{Q\left( {{\bf{G}},{\bf{v}},{t_S}} \right)}}\\
&\text{s.t.}\quad \mathop {\min }\limits_{\Delta {h_{DS}},{\bf{\Delta }}{{\bf{h}}_{DT}},{\bf{\Delta }}{{\bf{h}}_{TS}}} {R_D}\left( {{\bf{G}},{\bf{v}}} \right) \ge \frac{{{t_D}}}{{{\alpha _D}}}, \\
 &\mathop {\min }\limits_{{\bf{\Delta }}{{\bf{h}}_{TS}},\Delta {h_{RS}}} {R_R}\left( {{\bf{G}},{\bf{v}}} \right) \ge \frac{{{t_R}}}{{{\alpha _R}}},\\
  &\mathop {\min }\limits_{{\bf{\Delta }}{{\bf{h}}_{TS}}} {R_M}\left( {{\bf{G}},{\bf{v}},{\bf{u}}} \right) \ge \mathop {\max }\limits_{\Delta {h_{DS}},{\bf{\Delta }}{{\bf{h}}_{DT}},{\bf{\Delta }}{{\bf{h}}_{TS}}} {R_D}\left( {{\bf{G}},{\bf{v}}} \right),\label{Robust_S_M}\\
   &\frac{{{t_R}}}{{{\alpha _R}}} \ge {R_{th}},\label{Robust_S_R}\\
   &{t_S} + \sigma _T^2\left\| {\bf{G}} \right\|_F^2 + {\left\| {\bf{v}} \right\|^2} \le {P_{\max }},\label{Robust_S_power}\\
      &\mathop {\max }\limits_{{\bf{\Delta }}{{\bf{h}}_{TS}}} {P_S}{\left\| {{\bf{G}}{{\bf{h}}_{TS}}} \right\|^2} \le {t_S},\\
&{\left\| {\bf{u}} \right\|^2} \le 1,\label{Robust_S_u}\\
&(\ref{ICSI_1}),(\ref{ICSI_2}),
\end{align}
\end{subequations}
where $Q\left( {{\bf{G}},{\bf{v}},{t_S}} \right) = {{( {{t_S} + \sigma _T^2\| {\bf{G}} \|_F^2 + {{\| {\bf{v}} \|}^2}} )} \mathord{\left/
 {\vphantom {{\left( {{t_S} + \sigma _T^2\left\| {\bf{G}} \right\|_F^2 + {{\left\| {\bf{v}} \right\|}^2}} \right)} \xi }} \right.
 \kern-\nulldelimiterspace} \xi } + {N_T}{P_A} + {N_R}{P_R} + {P_C}$.

\begin{remark}
The constraint in (\ref{Robust_u}) can be replaced by (\ref{Robust_S_u}) without changing the optimal value of problem (\ref{Robust_Simplify}).
\end{remark}
\begin{proof}
Suppose that ${{{\bf{u}}^*}}$ is the optimal solution of problem (\ref{Robust_Simplify}) with ${\left\| {{{\bf{u}}^*}} \right\|^2} < 1$. Since ${\bf{u}}$ is only related to ${R_M}\left( {{\bf{G}},{\bf{v}},{\bf{u}}} \right)$,  we can always find another solution ${{\bf{\bar u}} = \alpha {{\bf{u}}^*},\alpha  > 1}$, which satisfies ${\left\| {{\bf{\bar u}}} \right\|^2} = 1$ and ${R_M}\left( {{\bf{G}},{\bf{v}},{\bf{\bar u}}} \right) = {R_M}\left( {{\bf{G}},{\bf{v}},{{\bf{u}}^*}} \right)$. Therefore, ${{\bf{\bar u}}}$ is also the optimal solution to problem (\ref{Robust_Simplify}) without changing the optimal value.
\end{proof}

The robust design in (\ref{Robust_Simplify}) is more challenging than
its counterpart for perfect CSIs, i.e., problem (\ref{pro_1}), and thus the
proposed iterative path-following algorithm is not applicable
due to the semi-infiniteness caused by CSI errors. To tackle this case, we subsequently develop an efficient AO method by combining Dinkelbach's algorithm \cite{dinkelbach1967nonlinear,cheung2013achieving} and weighted minimum mean square error (WMMSE) algorithm \cite{shi2011iteratively,gong2017energy}.

\subsection{Proposed AO Method}
Here, we first utilize the main idea of WMMSE algorithm \cite{shi2011iteratively,gong2017energy} to problem (\ref{Robust_Simplify}), which can transform the achievable rate maximization problem into a WMMSE problem by introducing
some auxiliary variables. We then apply the sign-definiteness lemma \cite{gharavol2012sign} to address the channel uncertainties and employ the structure of Dinkelbach's algorithm \cite{dinkelbach1967nonlinear,cheung2013achieving} to combat the nonlinear fractional objective function. Next, we propose an efficient AO method to solve problem (\ref{Robust_Simplify}). To this end, we first introduce the following lemmas.
\begin{lemma}\label{Rob_1}
(\cite{shi2011iteratively}): Define the mean square error (MSE) as
\begin{align}
M \buildrel \Delta \over = \left( {DB - 1} \right){\left( {DB - 1} \right)^H} + DR{D^H},
\end{align}
where $R > 0$. Then, we have
\begin{align}
\ln (1 + {B^H}{R^{ - 1}}B) = \mathop {\max }\limits_{S > 0,D} \ln S - SM + 1.
\end{align}
\end{lemma}
\begin{lemma}\label{Rob_2}
(\cite{shi2011iteratively}): Let $S$ be a positive scalar, and we have
\begin{align}
 - \ln T = \mathop {\max }\limits_{S > 0} \ln S - ST + 1.
\end{align}
\end{lemma}

Now, we define the MMSE linear equalizer ${D_D}$ for suspicious destination D to recover the transmitted signal ${x_S}$. The MSE ${M_D}$ at D can then be expressed as
\begin{equation*}
{M_D} =
\begin{cases}
{P_D}P_D^H + {D_D}{J_D}\left( {{\bf{G}},{\bf{v}}} \right)D_D^H, &\text{for NNPD},\\
{{\bar P}_D}\bar P_D^H + {D_D}{{\bar J}_D}\left( {{\bf{G}},{\bf{v}}} \right)D_D^H, &\text{for NPD},
\end{cases}
\end{equation*}
where ${{P_D} = \sqrt {{P_S}} {D_D}{h_{DS}} - 1}$ and ${{{\bar P}_D} = \sqrt {{P_S}} {D_D}{{\bar h}_{DS}}\left( {\bf{G}} \right) - 1}$, respectively.

Applying \textbf{Lemma \ref{Rob_1}}, we have
\begin{align}\label{Robust_C1}
{R_D}\left( {{\bf{G}},{\bf{v}}} \right) = \mathop {\max }\limits_{{S_D} > 0,{D_D}} \ln {S_D} - {S_D}{M_D} + 1.
\end{align}
Similarly, we have
\begin{align}
{R_R}\left( {{\bf{G}},{\bf{v}}} \right) &= \mathop {\max }\limits_{{S_R} > 0,{D_R}} \ln {S_R} - {S_R}{M_R} + 1,\\
{R_M}\left( {{\bf{G}},{\bf{v}},{\bf{u}}} \right) &= \mathop {\max }\limits_{{S_M} > 0,{D_M}} \ln {S_M} - {S_M}{M_M} + 1,
\end{align}
where ${M_R} = \left( {{D_R}{{\bf{h}}_{RT}}{\bf{v}} - 1} \right){\left( {{D_R}{{\bf{h}}_{RT}}{\bf{v}} - 1} \right)^H} + {D_R}{J_R}\left( {\bf{G}} \right)D_R^H$, ${M_M} = {P_M}P_M^H + {D_M}{J_M}\left( {{\bf{G}},{\bf{v}},{\bf{u}}} \right)D_M^H$ and ${{P_M} = \sqrt {{P_S}} {D_M}{{\bf{u}}^H}{{\bf{H}}_{MT}}{{\bf{V}}_{\bf{0}}}{\bf{G}}{{\bf{h}}_{TS}} - 1}$.

For the right-hand side of constraint (\ref{Robust_S_M}), we first rewrite ${R_D}\left( {{\bf{G}},{\bf{v}}} \right)$ as
\begin{align}
{R_D}\left( {{\bf{G}},{\bf{v}}} \right)={{R_{D,1}}\left( {{\bf{G}},{\bf{v}}} \right)}-{{R_{D,2}}\left( {{\bf{G}},{\bf{v}}} \right)},
\end{align}
where ${{R_{D,1}}\left( {{\bf{G}},{\bf{v}}} \right)} \buildrel \Delta \over = {\ln {M_t}}$ with
\begin{equation*}
{M_t} =
\begin{cases}
 \sigma _D^{ - 2}( {{J_D}( {{\bf{G}},{\bf{v}}} ) + {P_S}{{| {{h_{DS}}} |}^2}} ), &\text{for NNPD},\\
\sigma _D^{ - 2}( {{{\bar J}_D}( {{\bf{G}},{\bf{v}}} )+ {P_S}{{| {{{\bar h}_{DS}}( {\bf{G}} )} |}^2}}), &\text{for NPD},
\end{cases}
\end{equation*}
and
\begin{equation*}
{{R_{D,2}}\left( {{\bf{G}},{\bf{v}}} \right)} \buildrel \Delta \over =
\begin{cases}
{\ln \left[ { \sigma _D^{ - 2}\left( {{J_D}\left( {{\bf{G}},{\bf{v}}} \right)} \right)} \right]}, &\text{for NNPD},\\
{\ln \left[ {\sigma _D^{ - 2}\left( {{{\bar J}_D}\left( {{\bf{G}},{\bf{v}}} \right)} \right)} \right]}, &\text{for NPD}.
\end{cases}
\end{equation*}

Employing \textbf{Lemma \ref{Rob_2}} and \textbf{\ref{Rob_1}}, we have
\begin{align}
 - {R_{D,1}}\left( {{\bf{G}},{\bf{v}}} \right) = \mathop {\max }\limits_{{S_t} > 0} \ln {S_t} - {S_t}{M_t} + 1,\\
 {R_{D,2}}\left( {{\bf{G}},{\bf{v}}} \right) = \mathop {\max }\limits_{{S_d} > 0,{{\bf{D}}_{\bf{d}}}} \ln {S_d} - {S_d}{M_d} + 1,\label{Robust_CL}
\end{align}
where
\begin{equation*}
{M_d} =
\begin{cases}
{{P}_{d}}{P}_{d}^H + \sigma _D^2{{\bf{D}}_{\bf{d}}}{\bf{D}}_{\bf{d}}^H, &\text{for NNPD},\\
{{{\bar P}}_{d}}{\bar P}_{d}^H + \sigma _D^2{{\bf{D}}_{\bf{d}}}{\bf{D}}_{\bf{d}}^H, &\text{for NPD},
\end{cases}
\end{equation*}
with ${{P}_{d}} = {{\bf{D}}_{\bf{d}}}{[\sqrt {{P_S}} {{\bf{h}}_{DT}}{{\bf{V}}_{\bf{0}}}{\bf{G}}{{\bf{h}}_{TS}},{\sigma _T}{({{\bf{h}}_{DT}}{{\bf{V}}_{\bf{0}}}{\bf{G}})^H},{{\bf{h}}_{DT}}{\bf{v}}]^T}-1$ and ${{{\bar P}}_{d}} = {{\bf{D}}_{\bf{d}}}{[{\sigma _T}{({{\bf{h}}_{DT}}{{\bf{V}}_{\bf{0}}}{\bf{G}})^H},{{\bf{h}}_{DT}}{\bf{v}}]^T} - 1$.

Inserting (\ref{Robust_C1})-(\ref{Robust_CL}) into problem (\ref{Robust_Simplify}), and employing the auxiliary variables $\mathcal{B}  = \left\{ {{\beta _D},{\beta _R},{\beta _M},{\beta _t},{\beta _d}} \right\}$, we can rewrite problem (\ref{Robust_Simplify}) as
\begin{subequations}\label{Robust_MMSE}
\begin{align}&\mathop {\max }\limits_{{\bf{G}},{\bf{v}},{\bf{u}},{t_D},{t_R},{t_S}\hfill\atop
\scriptstyle\mathcal{S},\mathcal{D},\mathcal{B}}\quad \mathop \frac{{{t_D} + {t_R}}}{{Q\left( {{\bf{G}},{\bf{v}},{t_S}} \right)}}\label{Robust_MMSE_obj}\\
&\text{s.t.}\quad \ln {S_D} - {\beta _D} + 1 \ge \frac{{{t_D}}}{{{\alpha _D}}},\\
 &\ln {S_R} - {\beta _R} + 1 \ge \frac{{{t_R}}}{{{\alpha _R}}},\\
  &\ln {S_M} - {\beta _M} + \ln {S_t} - {\beta _t} + \ln {S_d} - {\beta _d} + 3 \ge 0,\\
 &{S_D}{M_D} \le {\beta _D},\forall \Delta {h_{DS}},{\bf{\Delta }}{{\bf{h}}_{DT}},{\bf{\Delta }}{{\bf{h}}_{TS}},\label{Robust_MMSE_CD}\\
&{S_R}{M_R} \le {\beta _R},\forall {\bf{\Delta }}{{\bf{h}}_{TS}},\Delta {h_{RS}},\label{Robust_MMSE_CR}\\
&{S_M}{M_M} \le {\beta _M},\forall {\bf{\Delta }}{{\bf{h}}_{TS}},\label{Robust_MMSE_CM}\\
&{S_t}{M_t} \le {\beta _t},\forall \Delta {h_{DS}},{\bf{\Delta }}{{\bf{h}}_{DT}},{\bf{\Delta }}{{\bf{h}}_{TS}},\label{Robust_MMSE_Ct}\\
&{S_d}{M_d} \le {\beta _d},\forall {\bf{\Delta }}{{\bf{h}}_{DT}},{\bf{\Delta }}{{\bf{h}}_{TS}},\label{Robust_MMSE_Cd}\\
&{P_S}{\left\| {{\bf{G}}{{\bf{h}}_{TS}}} \right\|^2} \le {t_S},\forall {\bf{\Delta }}{{\bf{h}}_{TS}},\label{Robust_MMSE_CS}\\
&(\ref{Robust_S_R}),(\ref{Robust_S_power}),(\ref{Robust_S_u}),(\ref{ICSI_1}),(\ref{ICSI_2}),
\end{align}
\end{subequations}
where $\mathcal{S} = \left\{ {{S_D},{S_R},{S_M},{S_t},{S_d}} \right\} > 0$ and $\mathcal{D} = \left\{ {{D_D},{D_R},{D_M},{{\bf{D}}_{\bf{d}}}} \right\}$, respectively.

Obviously, the nonlinear fractional objective function in problem (\ref{Robust_MMSE}) is neither convex nor concave, and thus standard convex optimization algorithms cannot be applied. Fortunately, according to fractional programming, Dinkelbach’s algorithm \cite{dinkelbach1967nonlinear,cheung2013achieving} can be utilized, in which we can transform the objective function (\ref{Robust_MMSE_obj}) into a parametric subtractive form given by
\begin{align}
F\left( {\lambda ,{\bf{G}},{\bf{v}},{t_S},{t_D},{t_R}} \right) = {t_D} + {t_R} - \lambda Q\left( {{\bf{G}},{\bf{v}},{t_S}} \right).
\end{align}
Specifically, by applying Dinkelbach’s algorithm \cite{dinkelbach1967nonlinear,cheung2013achieving}, we confirm that the parameter $\lambda$ that causes zero output of the corresponding parametric subtractive function $F\left( {\lambda ,{\bf{G}},{\bf{v}},{t_S},{t_D},{t_R}} \right)$ is the optimal solution of problem (\ref{Robust_MMSE}).
However, problem (\ref{Robust_MMSE}) remains intractable due to the semi-infinite constraints. In the following, we focus on these semi-infinite constraints and attempt to eliminate them.

First, we focus on constraint (\ref{Robust_MMSE_CR}). To proceed, we rewrite the term ${S_R}{M_R}$ in constraint (\ref{Robust_MMSE_CR}) as
\begin{align}\label{Ve_R}
{S_R}{M_R} = {\left\| {{\bm{\phi} _R}} \right\|^2} = {\left\| \left[
 \begin{matrix}
   {E_R}\left( {{D_R}{{\bf{h}}_{RT}}{\bf{v}} - 1} \right) \\
   \sqrt {{P_S}} {E_R}{D_R}{{\bf{h}}_{RT}}{{\bf{V}}_{\bf{0}}}{\bf{G}}{{\bf{h}}_{TS}} \\
   {\sigma _T}{E_R}{D_R}{\left( {{{\bf{h}}_{RT}}{{\bf{V}}_{\bf{0}}}{\bf{G}}} \right)^H} \\
   \sqrt {{P_S}} {E_R}{D_R}{h_{RS}} \\
   {\sigma _R}{E_R}{D_R}
  \end{matrix}
  \right] \right\|^2},
\end{align}
where ${E_R} = \sqrt {{S_R}}$. After applying (\ref{ICSI_TS}) and (\ref{ICSI_RS}) into (\ref{Ve_R}), we have
\begin{align}
{\bm{\phi} _R} \buildrel \Delta \over = {\bm{\hat \phi }_R} + {\bm{\Delta} _R},
\end{align}
where ${\bm{\Delta} _R} \buildrel \Delta \over = \bm{\Omega} _{TS}^R{\bf{\Delta }}{{\bf{h}}_{TS}} + \bm{\Omega} _{RS}^R\Delta {h_{RS}}$,
${\bm{\hat \phi }_R} = \left[
 \begin{matrix}
   {E_R}\left( {{D_R}{{\bf{h}}_{RT}}{\bf{v}} - 1} \right) \\
   \sqrt {{P_S}} {E_R}{D_R}{{\bf{h}}_{RT}}{{\bf{V}}_{\bf{0}}}{\bf{G}}{{{\bf{\hat h}}}_{TS}} \\
   {\sigma _T}{E_R}{D_R}{\left( {{{\bf{h}}_{RT}}{{\bf{V}}_{\bf{0}}}{\bf{G}}} \right)^H} \\
   \sqrt {{P_S}} {E_R}{D_R}{{\hat h}_{RS}} \\
   {\sigma _R}{E_R}{D_R}
  \end{matrix}
  \right]$,
$\bm{\Omega} _{TS}^R = {[0,\sqrt {{P_S}} {E_R}{D_R}{{\bf{h}}_{RT}}{{\bf{V}}_{\bf{0}}}{\bf{G}},{\bf{0}},0,0]^T}$ and $\bm{\Omega} _{RS}^R = {[0,0,{\bf{0}},\sqrt {{P_S}} {E_R}{D_R},0]^T}$.

By employing the Schur complement lemma \cite{schott2016matrix}, the constraint
(\ref{Robust_MMSE_CR}) is rewritten as
\begin{align}\label{SCH_R}
\left[
 \begin{matrix}
   {{\beta _R}} & {\bm{\hat \phi} _R^H} \\
   {{{\bm{\hat \phi} }_R}} & {\bf{I}}
  \end{matrix}
  \right] \succeq  - \left[
 \begin{matrix}
   {\bf{0}} & {\bm{\Delta} _R^H} \\
   {{\bm{\Delta} _R}} & {\bf{0}}
  \end{matrix}
  \right].
\end{align}
The constraint (\ref{SCH_R}) still contains channel uncertainties. To eliminate these uncertainties, we need the following sign-definiteness lemma.
\begin{lemma}\label{Rob_3}
(\cite{gharavol2012sign}): Given a Hermitian matrix ${\bf{A}}$ and arbitrary matrices $\left\{ {{{\bf{P}}_i},{{\bf{Q}}_i}} \right\}_{i = 1}^N$, the semi-infinite linear matrix inequality (LMI) of the form
\begin{align}
{\bf{A}} \succeq \sum\limits_{i = 1}^N {\left( {{\bf{P}}_i^H{{\bf{X}}_i}{{\bf{Q}}_i} + {\bf{Q}}_i^H{\bf{X}}_i^H{{\bf{P}}_i}} \right)} ,\forall {{\bf{X}}_i}:\left\| {{{\bf{X}}_i}} \right\| \le {\epsilon _i}
\end{align}
holds if and only if there exist non-negative real numbers ${\lambda _1}, \cdots {\lambda _N}$ such that
\begin{align}
\left[
 \begin{matrix}
   {{\bf{A}} - \sum\limits_{i = 1}^N {{\lambda _i}{\bf{Q}}_i^H{{\bf{Q}}_i}} } & { - {\epsilon _1}{\bf{P}}_1^H} & \cdots & { - {\epsilon _N}{\bf{P}}_N^H} \\
   { - {\epsilon _1}{{\bf{P}}_1}} & {{\lambda _1}{\bf{I}}} & \cdots & {\bf{0}} \\
   \vdots & \vdots & \ddots & \vdots \\
   { - {\epsilon _N}{{\bf{P}}_N}} & {\bf{0}} & \cdots & {{\lambda _N}{\bf{I}}}
  \end{matrix}
  \right] \succeq {\bf{0}}.
\end{align}
\end{lemma}

According to \textbf{Lemma \ref{Rob_3}}, we properly choose its parameters as follows
\begin{align}
{{\bf{A}}_R} = \left[
 \begin{matrix}
   {{\beta _R}} & {\bm{\hat \phi} _R^H} \\
   {{{\bm{\hat \phi} }_R}} & {\bf{I}}
  \end{matrix}
  \right],\ &{\bf{Q}}_{TS}^R = {\bf{Q}}_{RS}^R = \left[ { - 1\;0} \right],\\
{\bf{P}}_{TS}^R = [ {{\bf{0}}\;{{( {\bm{\Omega} _{TS}^R} )}^H}} ],\ &{\bf{P}}_{RS}^R = [ {0\;{{( {\bm{\Omega} _{RS}^R} )}^H}} ].
\end{align}
Therefore, the constraint (\ref{Robust_MMSE_CR}) can be further reduced to an LMI given by
\begin{align}\label{PD_SR}
\left[
 \begin{matrix}
   \left[
 \begin{matrix}
   {{\beta _R} - {\mu _{TS}} - {\mu _{RS}}} & {\bm{\hat \phi} _R^H} \\
   {{\bm{\hat \phi }_R}} & {\bf{I}}
  \end{matrix}
  \right] & {{\bm{\theta }}_R^H} \\
   {{{\bm{\theta }}_R}} & {\text{diag}\{ {\mu _{TS}}{\bf{I}},{\mu _{RS}}{\bf{I}}\} }
  \end{matrix}
  \right] \succeq {\bf{0}},
\end{align}
where ${{\bm{\theta }}_R} =  - {[ {{\epsilon _{TS}}{{( {{\bf{P}}_{TS}^R} )}^H},{\epsilon _{RS}}{{( {{\bf{P}}_{RS}^R} )}^H}} ]^H}$.

Similarly, we can rewrite (\ref{Robust_MMSE_CM}) and (\ref{Robust_MMSE_CS}) as the following LMIs:
\begin{align}
\left[
 \begin{matrix}
   \left[
 \begin{matrix}
   {{\beta _M} - {\sigma _{TS}}} & {\bm{\hat \phi} _M^H} \\
   {{\bm{\hat \phi }_M}} & {\bf{I}}
  \end{matrix}
  \right] & {{\bm{\theta }}_M^H} \\
   {{{\bm{\theta }}_M}} & {\text{diag}\{ {\sigma _{TS}}{\bf{I}}\} }
  \end{matrix}
  \right]\succeq {\bf{0}},\label{PD_M}\\
\left[
 \begin{matrix}
   \left[
 \begin{matrix}
   {{t_S} - {\tau _{TS}}} & {\bm{\hat \phi} _S^H} \\
   {{\bm{\hat \phi }_S}} & {\bf{I}}
  \end{matrix}
  \right] & {{\bm{\theta }}_S^H} \\
   {{{\bm{\theta }}_S}} & {\text{diag}\{ {\tau _{TS}}{\bf{I}}\} }
  \end{matrix}
  \right] \succeq {\bf{0}},\label{PD_POWER}
\end{align}
where ${\bm{\hat \phi }_M} = \left[
 \begin{matrix}
   {E_M}( {{D_M}\sqrt {{P_S}} {{\bf{u}}^H}{{\bf{H}}_{MT}}{{\bf{V}}_{\bf{0}}}{\bf{G}}{{{\bf{\hat h}}}_{TS}} - 1} ) \\
   {\sigma _T}{E_M}{D_M}{( {{{\bf{u}}^H}{{\bf{H}}_{MT}}{{\bf{V}}_{\bf{0}}}{\bf{G}}} )^H} \\
   {E_M}{D_M}{{\bf{u}}^H}{{\bf{H}}_{MT}}{\bf{v}} \\
   {\sigma _M}{E_M}{D_M}{\bf{u}}
  \end{matrix}
  \right]$,
${E_M} = \sqrt {{S_M}}$,
${{\bm{\theta }}_M} =  - {\epsilon _{TS}}[ {{\bf{0}}\;{{( {\bm{\Omega} _{TS}^M} )}^H}} ]$, ${\bm{\hat \phi }_S} = \sqrt {{P_S}} {\bf{G}}{{{\bf{\hat h}}}_{TS}}$, ${{\bm{\theta }}_S} =  - {\epsilon _{TS}}\left[ {{\bf{0}}\;\sqrt {{P_S}} {{\bf{G}}^H}} \right]$ and $\bm{\Omega} _{TS}^M = {[\sqrt {{P_S}} {E_M}{D_M}{{\bf{u}}^H}{{\bf{H}}_{MT}}{{\bf{V}}_{\bf{0}}}{\bf{G}},{\bf{0}},0,{\bf{0}}]^T}$.

In a similar manner, we can transform (\ref{Robust_MMSE_CD}), (\ref{Robust_MMSE_Ct}), and (\ref{Robust_MMSE_Cd}) to the following, respectively, LMIs for NNPD,
\begin{align}\label{PD_susis_D}
\left[
 \begin{matrix}
   {{{\bf{T}}_D}} & {{\bm{\theta }}_D^H} \\
   {{{\bm{\theta }}_D}} & {\text{diag}\{ {\lambda _{DS}}{\bf{I}},{\lambda _{DT}}{\bf{I}},{\lambda _{TS}}{\bf{I}}\} }
  \end{matrix}
  \right] \succeq {\bf{0}},
\end{align}
\begin{align}\label{PD_ALTER}
\left[
 \begin{matrix}
   {{{\bf{T}}_t}} & {{\bm{\theta }}_t^H} \\
   {{{\bm{\theta }}_t}} & {\text{diag}\{ {\eta _{DT}}{\bf{I}},{\eta _{TS}}{\bf{I}},{\eta _{DS}}{\bf{I}}\} }
  \end{matrix}
  \right] \succeq {\bf{0}},
\end{align}
and
\begin{align}\label{PD_susis_Dd}
 \left[
 \begin{matrix}
    \left[
 \begin{matrix}
   {{\beta _d} - {\nu _{DT}} - {\nu _{TS}}} & {\bm{\hat \phi} _d^H} \\
   {{\bm{\hat \phi }_d}} & {\bf{I}}
  \end{matrix}
  \right] & {{\bm{\theta }}_d^H}  \\
   {{{\bm{\theta }}_d}} & {\text{diag}\{ {\nu _{DT}}{\bf{I}},{\nu _{TS}}{\bf{I}}\} }
  \end{matrix}
  \right] \succeq {\bf{0}}.
\end{align}
Whereas for NPD, constraints (\ref{Robust_MMSE_CD}), (\ref{Robust_MMSE_Ct}), and (\ref{Robust_MMSE_Cd}) can be, respectively, expressed as
\begin{align}\label{PD_susis_D_NPD}
\left[
 \begin{matrix}
   {{{{\bf{\bar T}}}_D}} & {{\bm{\bar \theta }}_D^H} \\
   {{{{\bm{\bar \theta }}}_D}} & {\text{diag}\{ {\lambda _{DS}}{\bf{I}},{\lambda _{DT}}{\bf{I}},{\lambda _{TS}}{\bf{I}}\} }
  \end{matrix}
  \right] \succeq {\bf{0}},
\end{align}
\begin{align}\label{PD_ALTER_NPD}
\left[
 \begin{matrix}
   {{{{\bf{\bar T}}}_t}} & {{\bm{\bar \theta }}_t^H} \\
   {{{{\bm{\bar \theta }}}_t}} & {\text{diag}\{ {\eta _{DT}}{\bf{I}},{\eta _{TS}}{\bf{I}},{\eta _{DS}}{\bf{I}}\} }
  \end{matrix}
  \right] \succeq {\bf{0}},
\end{align}
and
\begin{align}\label{PD_susis_Dd_NPD}
\left[
 \begin{matrix}
    \left[
 \begin{matrix}
   {{\beta _d} - {\nu _{DT}}} & \hat{\bar{\bm{\phi}}}_d^H \\
   \hat{\bar{\bm{\phi}}}_d & {\bf{I}}
  \end{matrix}
  \right] & {{\bm{\bar \theta }}_d^H} \\
   {{{{\bm{\bar \theta }}}_d}} & {\text{diag}\{ {\nu _{DT}}{\bf{I}}\} }
  \end{matrix}
  \right] \succeq {\bf{0}}.
\end{align}
The details of the derivations for (\ref{PD_susis_D})-(\ref{PD_susis_Dd_NPD}) are provided in the Appendix.

On the basis of the above results, we can rewrite problem (\ref{Robust_MMSE}) as
\begin{subequations}\label{Robust_NNPD}
\begin{align}&\mathop {\max }\limits_{{\bf{G}},{\bf{v}},{\bf{u}},{t_D},{t_R},{t_S}\hfill\atop
\scriptstyle\lambda, \bf{E},\mathcal{D},\mathcal{B},{\bf{\chi }}}\quad \mathop F\left( {\lambda ,{\bf{G}},{\bf{v}},{t_S},{t_D},{t_R}} \right)\\
&\text{s.t.}\quad 2\ln {E_D} - {\beta _D} + 1 \ge \frac{{{t_D}}}{{{\alpha _D}}},\\
 &2\ln {E_R} - {\beta _R} + 1 \ge \frac{{{t_R}}}{{{\alpha _R}}},\\
  &2\ln {E_M} - {\beta _M} + 2\ln {E_t} - {\beta _t} + 2\ln {E_d} - {\beta _d} + 3 \ge 0,\\
  &(\ref{Robust_S_R}),(\ref{Robust_S_power}),(\ref{Robust_S_u}),(\ref{PD_SR}),(\ref{PD_M}),(\ref{PD_POWER}),(\ref{PD_susis_D}),(\ref{PD_ALTER}),(\ref{PD_susis_Dd})
\end{align}
\end{subequations}
for NNPD and
\begin{subequations}\label{Robust_NPD}
\begin{align}&\mathop {\max }\limits_{{\bf{G}},{\bf{v}},{\bf{u}},{t_D},{t_R},{t_S}\hfill\atop
\scriptstyle\lambda, \bf{E},\mathcal{D},\mathcal{B},{{\bf{\bar \chi }}}}\quad \mathop F\left( {\lambda ,{\bf{G}},{\bf{v}},{t_S},{t_D},{t_R}} \right)\\
&\text{s.t.}\quad 2\ln {E_D} - {\beta _D} + 1 \ge \frac{{{t_D}}}{{{\alpha _D}}},\\
 &2\ln {E_R} - {\beta _R} + 1 \ge \frac{{{t_R}}}{{{\alpha _R}}},\\
  &2\ln {E_M} - {\beta _M} + 2\ln {E_t} - {\beta _t} + 2\ln {E_d} - {\beta _d} + 3 \ge 0,\\
  &(\ref{Robust_S_R}),(\ref{Robust_S_power}),(\ref{Robust_S_u}),(\ref{PD_SR}),(\ref{PD_M}),(\ref{PD_POWER}),(\ref{PD_susis_D_NPD}),(\ref{PD_ALTER_NPD}),(\ref{PD_susis_Dd_NPD})
\end{align}
\end{subequations}
for NPD, respectively. In which ${\bf{E}} = \left\{ {{E_D},{E_R},{E_M},{E_t},{E_d}} \right\}$, and non-negative slack variables ${\bf{\bar \chi }} = \{ {\mu _{TS}},{\mu _{RS}},{\sigma _{TS}},{\tau _{TS}},{\lambda _{DS}},{\lambda _{DT}},{\lambda _{TS}},{\eta _{DT}},{\eta _{TS}},{\eta _{DS}},{\nu _{DT}}\}$ and ${\bf{\chi }} = \{ {\bf{\bar \chi }},{\nu _{TS}}\}$. Next, we introduce the proposed AO method to solve problem (\ref{Robust_NNPD}) for NNPD. Note that similar steps can be used to handle problem (\ref{Robust_NPD}) for NPD, and thus are omitted for brevity.

Obviously, problem (\ref{Robust_NNPD}) is no longer semi-infinite, but it remains non-convex due to the coupled optimization variables. As a consequence, we adopt AO method to optimize the variables $\lambda$, ${\bf{E}}$, $\mathcal{D}$, ${\bf{u}}$, and $\left\{ {{\bf{G}},{\bf{v}}} \right\}$ iteratively.
Specifically, $\lambda$ is updated in the outer layer. For the inner
layer, we optimize variables ${\bf{E}}$, $\mathcal{D}$, ${\bf{u}}$ and $\left\{ {{\bf{G}},{\bf{v}}} \right\}$ alternately, i.e., by fixing a subset of optimization variables, problem (\ref{Robust_NNPD}) reduces to a convex problem in the remaining variables, and thus can be efficiently solved by standard convex optimization techniques.

We can also simplify problem (\ref{Robust_NNPD}) as follows with respect to ${\bf{u}}$  when the other variables are fixed.
\begin{subequations}\label{Robust_U}
\begin{align}\mathop {\min }\limits_{{\beta _M},{{\sigma _{TS}}},{\bf{u}}}\quad &\mathop {{\beta _M}}\\
\text{s.t.}\quad & (\ref{PD_M}),(\ref{Robust_S_u}).
\end{align}
\end{subequations}

Based on these analyses, an iterative sequence of optimizations is performed until the desired convergence is reached. The detailed iterative AO method is presented in \textbf{Algorithm \ref{alg:B}}.
\begin{algorithm}[H]
\caption{Proposed AO Method to Solve Problem (\ref{Robust_NNPD})}
\label{alg:B}
\begin{algorithmic}
\STATE {Initializing: ${{\lambda ^{(0)}} = 0}$, $q: = 0$.}
\STATE {Employ \textbf{Algorithm \ref{alg:C}} to generate a feasible initial point $\left( {{{\bf{E}}^{(0)}},{\mathcal{D}^{(0)}},{{\bf{u}}^{(0)}},{{\bf{G}}^{(0)}},{{\bf{v}}^{(0)}}} \right)$.}
\REPEAT
\STATE \begin{itemize}
        \item [1] $\ell: = 0$.
        \item [2] \textbf{repeat}
        \begin{enumerate}
        \item Given fixed ${{\mathcal{D}^{(\ell)}}}$, ${{{\bf{u}}^{(\ell)}}}$, ${{{\bf{G}}^{(\ell)}}}$ and ${{{\bf{v}}^{(\ell)}}}$, solve (\ref{Robust_NNPD}) to obtain ${{{\bf{E}}^{(\ell + 1)}}}$;
        \item Given fixed ${{{\bf{E}}^{(\ell + 1)}}}$, ${{{\bf{u}}^{(\ell)}}}$, ${{{\bf{G}}^{(\ell)}}}$ and ${{{\bf{v}}^{(\ell)}}}$, solve (\ref{Robust_NNPD}) to obtain ${{\mathcal{D}^{(\ell+1)}}}$;
        \item Given fixed ${{{\bf{E}}^{(\ell + 1)}}}$, ${{\mathcal{D}^{(\ell+1)}}}$, ${{{\bf{G}}^{(\ell)}}}$ and ${{{\bf{v}}^{(\ell)}}}$, solve (\ref{Robust_U}) to obtain ${{{\bf{u}}^{(\ell+1)}}}$;
        \item Given fixed ${{{\bf{E}}^{(\ell + 1)}}}$, ${{\mathcal{D}^{(\ell+1)}}}$ and ${{{\bf{u}}^{(\ell+1)}}}$, solve (\ref{Robust_NNPD}) to obtain ${{{\bf{G}}^{(\ell+1)}}}$ and ${{{\bf{v}}^{(\ell+1)}}}$;
         \item $\ell: = \ell + 1$;
        \end{enumerate}
         \item [3] \textbf{until} convergence of the objective value in (\ref{Robust_NNPD}).
         \item [4] Update ${{\bf{{\bar G}}}^{(q + 1)}}={{\bf{G}}^{(\ell+1)}},{{\bf{{\bar v}}}^{(q + 1)}}={{\bf{v}}^{(\ell+1)}},{\bar t}_S^{(q + 1)}=t_S^{(\ell+1)},{\bar t}_D^{(q + 1)}=t_D^{(\ell+1)}$ and ${\bar t}_R^{(q + 1)}=t_R^{(\ell+1)}$;
         \item [5] Update ${\lambda ^{(q + 1)}} = \frac{{{\bar t}_D^{(q + 1)} + {\bar t}_R^{(q + 1)}}}{{Q\left( {{{\bf{{\bar G}}}^{(q + 1)}},{{\bf{{\bar v}}}^{(q + 1)}},{\bar t}_S^{(q + 1)}} \right)}}$;
         \item [6] $q: = q + 1$;
       \end{itemize}
\UNTIL{$\left| {{\lambda ^{(q + 1)}} - {\lambda ^{(q)}}} \right| \le \iota$, where $\iota  > 0$ is a small value to ensure convergence.}
\end{algorithmic}
\end{algorithm}

\textsl{Initialization of \textbf{Algorithm \ref{alg:B}}:} To find a feasible initial point for problem (\ref{Robust_NNPD}), we consider the following optimization:
\begin{subequations}\label{Initial_A2}
\begin{align}&\mathop {\max }\limits_{{\bf{G}},{\bf{v}},{\bf{u}},{t},{t_S},\bf{E},\mathcal{D},\mathcal{B},{\bf{\chi }}}\quad \mathop t\\
&\text{s.t.}\quad 2\ln {E_D} - {\beta _D} + 1 \ge \frac{t}{{{\alpha _D}}},\\
&2\ln {E_M} - {\beta _M} + 2\ln {E_t} - {\beta _t} + 2\ln {E_d} - {\beta _d} + 3 \ge t,\\
&2\ln {E_R} - {\beta _R} + 1 - {R_{th}} \ge t,\\
&(\ref{Robust_S_power}),(\ref{Robust_S_u}),(\ref{PD_SR}),(\ref{PD_M}),(\ref{PD_POWER}),(\ref{PD_susis_D}),(\ref{PD_ALTER}),(\ref{PD_susis_Dd}).
\end{align}
\end{subequations}
A feasible initial point to execute \textbf{Algorithm \ref{alg:B}} can be the solution of problem (\ref{Initial_A2}) whenever $t \ge 0$. Here, we design \textbf{Algorithm \ref{alg:C}} to obtain a feasible initial point for \textbf{Algorithm \ref{alg:B}}.

\subsection{Convergence and Complexity Analysis}

\textsl{Convergence Analysis:}
Given ${\lambda ^{(q)}}$, we employ the proposed AO method to transform problem (\ref{Robust_NNPD})/(\ref{Robust_NPD}) into a series of subproblems, and optimize the variables alternatively in inner layer optimization. According to \cite{shi2011iteratively}, the inner layer optimization of AO method is guaranteed to converge to a locally optimal point of problem (\ref{Robust_NNPD})/(\ref{Robust_NPD}), where we have
\begin{align}
F\left( {{\lambda ^{(q)}},{{\bf{\Theta }}^{(q + 1)}}} \right) \ge F\left( {{\lambda ^{(q)}},{{\bf{\Theta }}^{(q)}}} \right),
\end{align}
with ${{\bf{\Theta }}^{(q)}} = \left\{ {{{\bf{{\bar G}}}^{(q)}},{{\bf{{\bar v}}}^{(q )}},{\bar t}_S^{(q )},{\bar t}_D^{(q )},{\bar t}_R^{(q )}} \right\}$.

For outer layer optimization, suppose that ${{\bf{\Theta }}^{(q)}}$ is a feasible point of inner layer optimization with $F\left( {{\lambda ^{(q)}},{{\bf{\Theta }}^{(q)}}} \right) = 0$, while ${{{\bf{\Theta }}^{(q + 1)}}}$ is the optimal solution of inner layer optimization. Therefore, $F\left( {{\lambda ^{(q)}},{{\bf{\Theta }}^{(q + 1)}}} \right) > 0$ as far as ${{\bf{\Theta }}^{(q + 1)}} \ne {{\bf{\Theta }}^{(q)}}$, which indicates
\begin{align}
{\lambda ^{(q)}} = &\frac{{{\bar t}_D^{(q )} + {\bar t}_R^{(q )}}}{{Q\left( {{{\bf{{\bar G}}}^{(q )}},{{\bf{{\bar v}}}^{(q)}},{\bar t}_S^{(q)}} \right)}}\nonumber\\
<& \frac{{{\bar t}_D^{(q + 1)} + {\bar t}_R^{(q + 1)}}}{{Q\left( {{{\bf{{\bar G}}}^{(q + 1)}},{{\bf{{\bar v}}}^{(q + 1)}},{\bar t}_S^{(q + 1)}} \right)}}= {\lambda ^{(q + 1)}}.
\end{align}
We also note that problem (\ref{Robust_Simplify}) is bounded due to the power limited constraint (\ref{Robust_S_power}). Consequently, we can conclude that the proposed AO method yields a monotonically increasing sequence of objective values for problem (\ref{Robust_Simplify}), and converges to a locally optimal solution.

\textsl{Complexity Analysis:}
The computational complexity of the proposed AO method is mainly dominated by inner layer optimization, which comprises four steps to solve the standard semidefinite programming (SDP) problem for both NNPD and NPD. In particular, the inner layer optimization problem for NNPD involves three SDP problems with three LMI constraints of size $2{N_R} + {N_T} + 6$, three LMI constraints of size $2{N_R}+ 6$, $2{N_R} + {N_M} + 3$ and ${N_T} + 1$, respectively, and one SDP problem with one LMI constraint of size $2{N_R} + {N_M} + 3$. Similarly, the inner layer optimization problem of NPD contains three SDP problems with two LMI constraints of dimension $2{N_R} + {N_T} + 5$, four LMI constraints of dimension $2{N_R} + 6$, $2{N_R} + {N_M} + 3$, ${N_T} + 1$ and ${N_T} + {N_R} + 3$, respectively, as well as one SDP problem with one LMI constraint of dimension $2{N_R} + {N_M} + 3$. From \cite{ben2001lectures}, the computational complexity for solving an SDP within an accuracy $\varepsilon $ is $\mathcal{O}\left( {(m{n^{3.5}} + {m^2}{n^{2.5}} + {m^3}{n^{0.5}})\log \left( {\frac{1}{\varepsilon }} \right)} \right)$, where $m$ is the number of semidefinite cone constraints and $n$ is the dimension of the semidefinite cone.

Based on the above analysis, we can approximately derive the per-iteration complexity order of the AO method in the inner layer optimization as (\ref{Compu_NNPD}) and (\ref{Compu_NPD}) for NNPD and NPD, respectively.
	\begin{figure*}[!t]
\begin{align}\label{Compu_NNPD}
{\mathcal{O}_{NNPD}} = \left[ {3\mathcal{O}\left( {3{{(2{N_R} + {N_T} + 6)}^{3.5}} + \bar N} \right) + \mathcal{O}\left( {{{(2{N_R} + {N_M} + 3)}^{3.5}}} \right)} \right]\log \left( {\frac{1}{\varepsilon }} \right),
\end{align}
\begin{align}\label{Compu_NPD}
{\mathcal{O}_{NPD}} = \left[ {3\mathcal{O}\left( {2{{(2{N_R} + {N_T} + 5)}^{3.5}} + \bar N + {{({N_T} + {N_R} + 3)}^{3.5}}} \right) + \mathcal{O}\left( {{{(2{N_R} + {N_M} + 3)}^{3.5}}} \right)} \right]\log \left( {\frac{1}{\varepsilon }} \right),
\end{align}
	\hrule
\end{figure*}
Here, we denote $\bar N = {(2{N_R} + 6)^{3.5}} + {(2{N_R} + {N_M} + 3)^{3.5}} + {({N_T} + 1)^{3.5}}$ for conciseness.
Finally, we can derive the computational complexity of the proposed AO method as ${n_o}{n_i}{\mathcal{O}_{NNPD}}$ and ${n_o}{n_i}{\mathcal{O}_{NPD}}$ for NNPD and NPD, respectively. Parameters ${n_o}$ and ${n_i}$ denote the
maximum outer and inner iteration numbers, respectively.

\section{Numerical Results}\label{numer}
\begin{figure}
  \centering
  \includegraphics[width=2.5in]{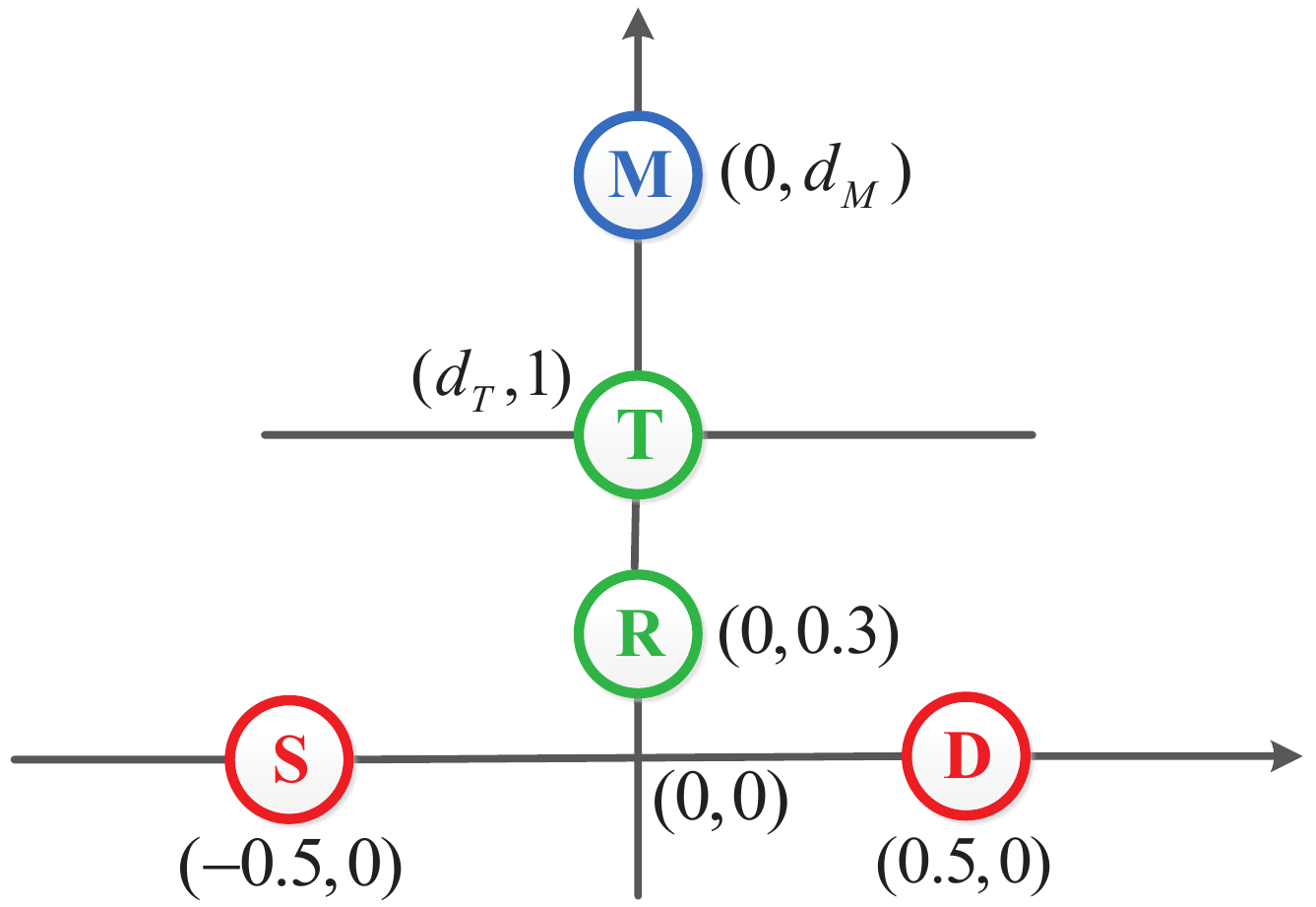}\\
  \caption{System topology.}\label{fig2}
\end{figure}
\begin{figure}
  \centering
  \includegraphics[width=3.2in]{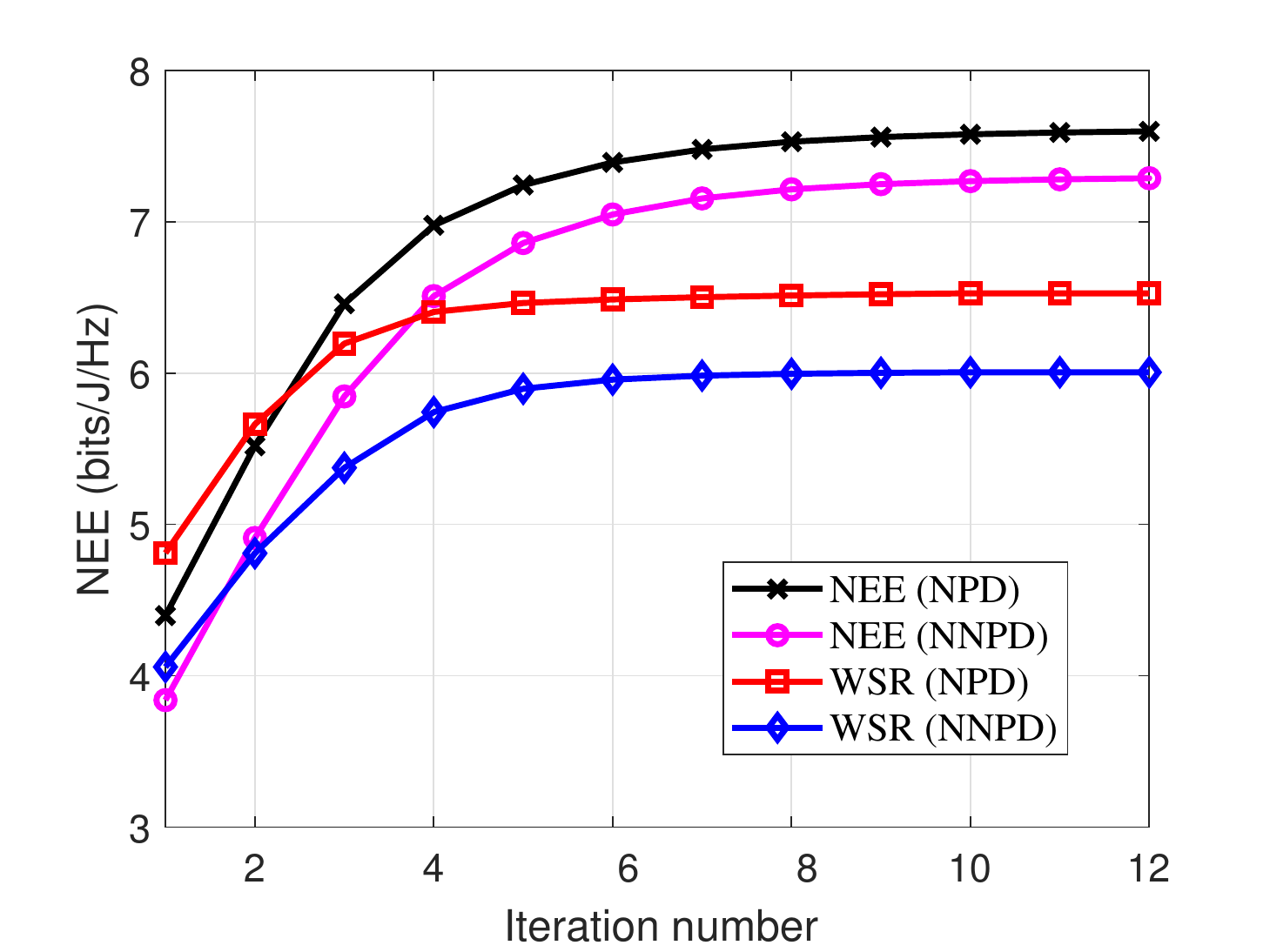}\\
  \caption{Achievable NEE versus iteration number of path-following algorithm.}\label{fig3}
\end{figure}
In this section, we present numerical simulations to evaluate the NEE performance achieved by our proposed schemes in both perfect and imperfect CSI cases. Unless otherwise specified, it is assumed that ${N_T} = 5$, ${N_R} = 3$, ${N_M} = 4$, ${P_S} = 10$ dBm, ${P_{\max }} = 25$ dBm, ${R_{th}} = 0.5$ nats/s/Hz, ${\alpha _D} = 1$, ${\alpha _R} = 1$, and the noise power $\sigma _T^2 = \sigma _D^2 = \sigma _R^2 = \sigma _M^2 = 1$ mW. The elements of all the channels involved are independent and identically distributed (i.i.d.) complex Gaussian random variables with zero mean. The variance of each entry of the channel responses is given by ${D_{ij}} = {D_0}d_{ij}^{ - \alpha }$, where ${D_0}$ is set to be 1, $\alpha $ is the path loss exponent set to be 3, and ${d_{ij}}$ denotes the distance from transmitter $j$ to receiver $i$. We normalize the distance between S and D as 1, and assume that the positions of S, D, T, R, and M are placed at $( - 0.5,0),(0.5,0),(0,1),(0,0.3)$, and $(0,2)$, respectively, as shown in Fig. \ref{fig2}. In addition, we set $\xi  = 40\%$, ${P_A} = 0.04$ W, ${P_R} = 0.02$ W, and ${P_C} = 0.05$ W \cite{xiong2012energy}. The standard
convex problem is conducted by CVX \cite{grant2014cvx}, and the number of channel realizations is averaged over 1000 independent Monte Carlo trials.

\subsection{Performance Evaluation of the Proposed Scheme
for Perfect CSI}
In this subsection, we consider the perfect CSI scenario and compare the NEE achieved by our proposed NEE maximization design with that achieved by weighted sum-rate (WSR) maximization design for both NNPD and NPD. Note that the WSR maximization design can be solved similarly by our path-following algorithm with minor modifications, and thus we omit the details here for brevity. To demonstrate the superiority, we also compare the performance of our proposed path-following algorithm to that of the Dinkelbach's algorithm combined with inner convex approximation (ICA) method (denoted as Dinkelbach-ICA), where the ICA is applied after the Dinkelbach's algorithm has been adopted.
\begin{algorithm}[H]
\caption{Initialization of \textbf{Algorithm \ref{alg:B}}}
\label{alg:C}
\begin{algorithmic}
\STATE {Initializing: $\ell: = 0$ and randomly generate ${\bf{G}}$, ${\bf{v}}$, ${\bf{u}}$, and $\mathcal{D}$.}
\REPEAT
\STATE \begin{enumerate}
         \item Solve problem (\ref{Initial_A2}) to update ${\bf{E}}$ with the other parameters fixed;
         \item Solve problem (\ref{Initial_A2}) to update $\mathcal{D}$ for fixed ${\bf{E}}$ found in the previous step;
         \item Solve problem (\ref{Initial_A2}) to update ${\bf{u}}$ for fixed $\mathcal{D}$ and ${\bf{E}}$ found in the previous steps;
         \item Solve problem (\ref{Initial_A2}) to update ${\bf{G}}$ and ${\bf{v}}$ for fixed ${\bf{u}}$, $\mathcal{D}$ and ${\bf{E}}$ found in the previous steps;
         \item $\ell: = \ell + 1$;
       \end{enumerate}
\UNTIL{${{t^{(\ell + 1)}} \ge 0}$.}
\STATE {Output: ${{{\bf{E}}^{(0)}} = {\bf{E}}}$, ${{\mathcal{D}^{(0)}} = \mathcal{D}}$, ${{{\bf{u}}^{(0)}} = {\bf{u}}}$, ${{{\bf{G}}^{(0)}} = {\bf{G}}}$ and ${{{\bf{v}}^{(0)}} = {\bf{v}}}$.}
\end{algorithmic}
\end{algorithm}

Fig. \ref{fig3} shows the achievable NEE performance obtained by our path-following algorithm versus the number of iterations under ${P_{\max }} = 25$ dBm with a randomly generated channel realization. It can be observed that the achievable NEE increases monotonically and converges within a certain number of iterations for all designs, which implies that our path-following algorithm can exhibit a good convergence property.

\begin{figure}
  \centering
  \includegraphics[width=3.2in]{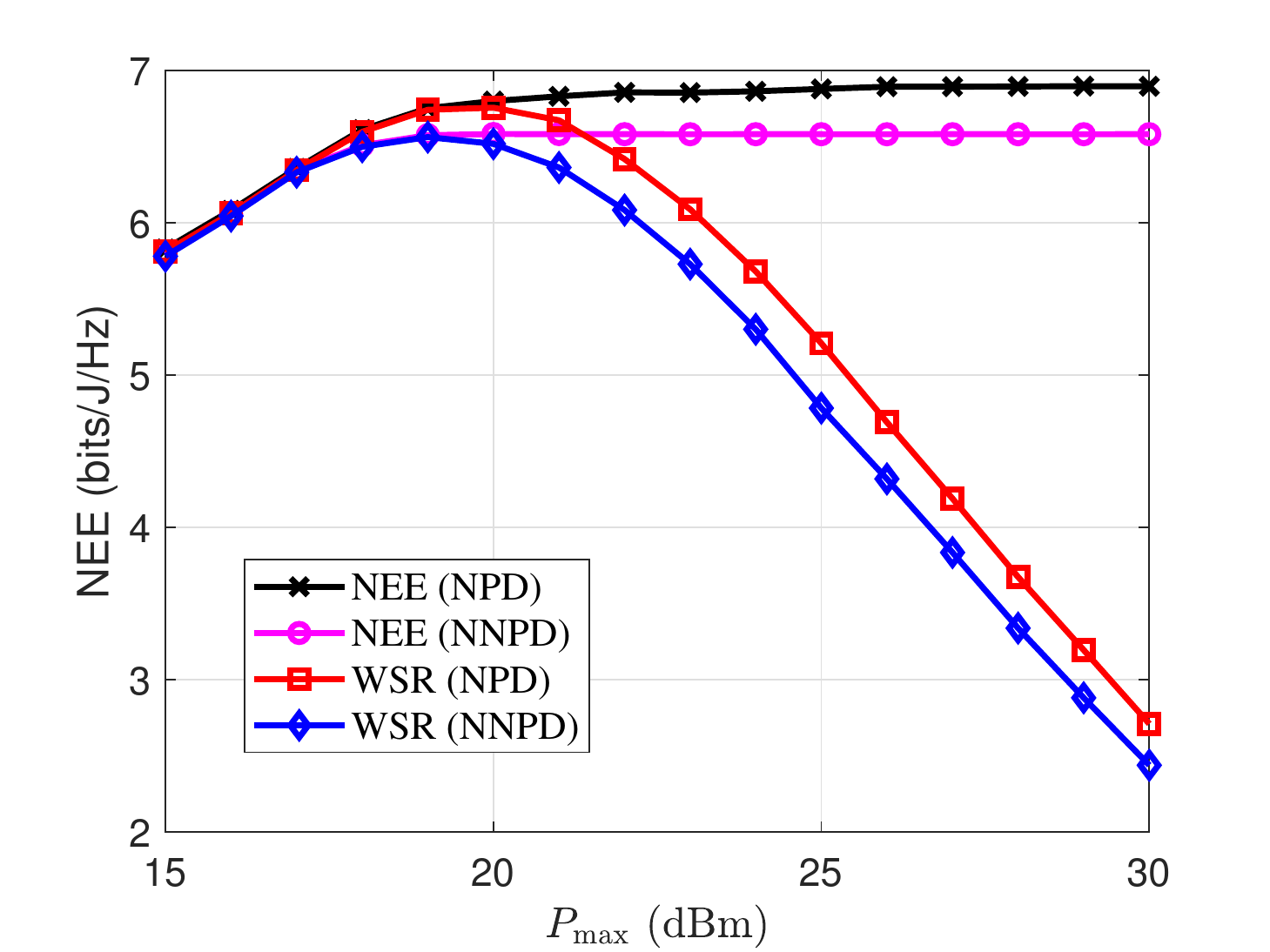}\\
\caption{Achievable NEE versus maximum available power ${P_{\max }}$.}\label{fig4}
\end{figure}

Fig. \ref{fig4} compares the achievable NEE performance versus maximum available power ${P_{\max }}$ at secondary transmitter T for different designs. We can observe that the proposed NEE maximization design outperforms the WSR maximization design for both NPD and NNPD cases, especially at high ${P_{\max }}$ regime. Specifically, the achievable NEE of the proposed NEE maximization design first increases and then reaches a saturation as ${P_{\max }}$ increases. The reason for this is mainly because our NEE maximization design can attain a good balance between energy efficiency and power consumption. However, the achievable NEE obtained by WSR maximization design first increases and then decreases with the increment of ${P_{\max }}$. This is due to the fact that, in order to maximize the WSR, the transmitter always radiates all the available power, which is not energy efficient. Moreover, it is noticed that when ${P_{\max }}$ is relatively small, the NEE maximization design and the WSR maximization design achieve similar NEE performance. The reason is that, when ${P_{\max }}$ is small, all available power is used in both designs. In this case, although the NEE maximization design aims to maximize the NEE, its power is fully utilized, leading to NEE performance that is similar to that of WSR maximization design.

We also notice that, with the increase of ${P_{\max }}$, the achievable NEE performance of NPD is superior to that of NNPD, although they have similar NEE performance when ${P_{\max }}$ is small. The reason lies in that the NPD at T can achieve a certain higher eavesdropping rate since it is able to spoof the suspicious link to increase its transmission rate by constructively forwarding the signal from S to D (probably occurs at high ${P_{\max }}$ regions), which is impossible for NNPD.

\begin{figure}
  \centering
  \includegraphics[width=3.2in]{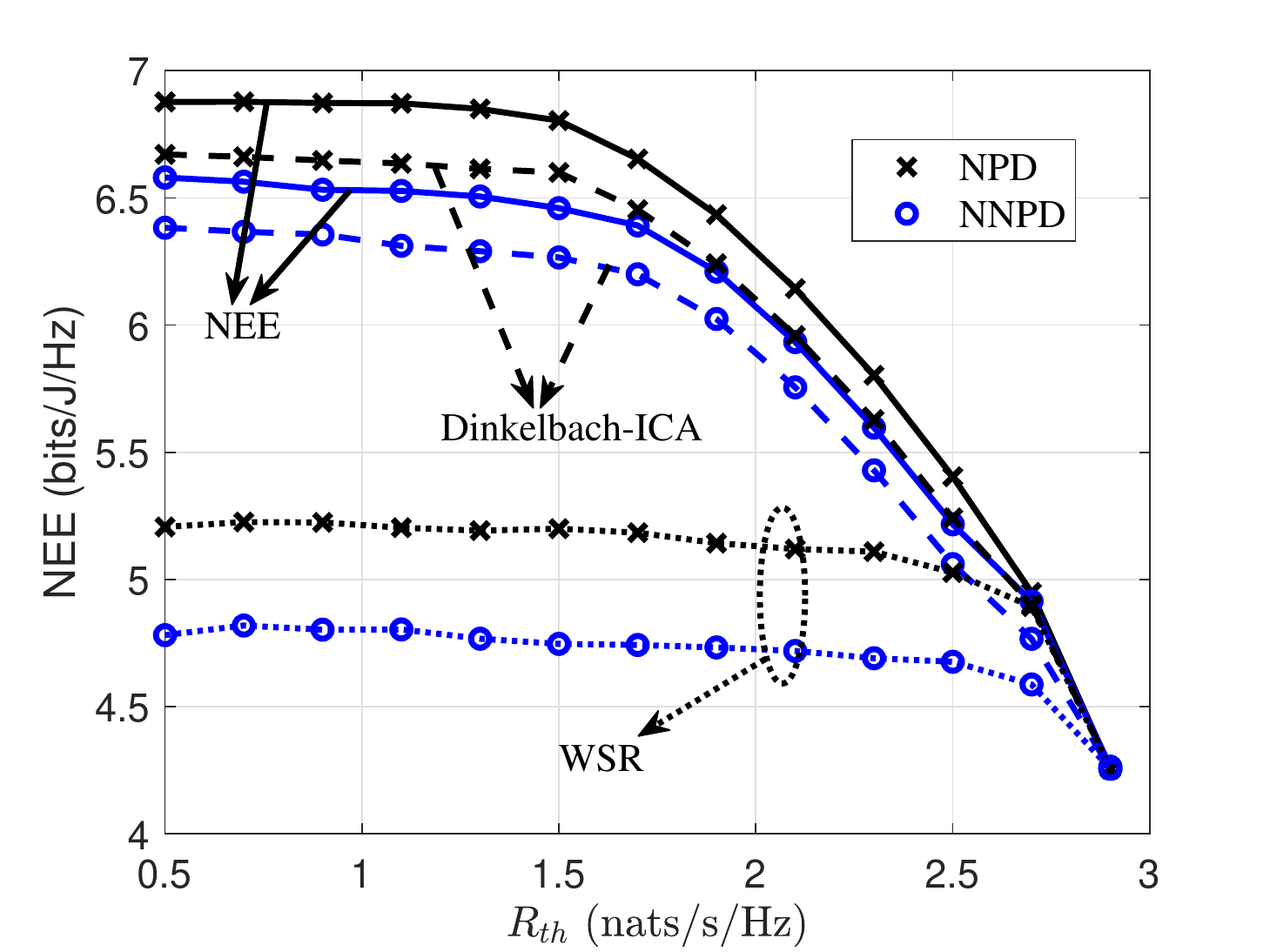}\\
\caption{Achievable NEE versus minimum data rate requirement ${R_{th}}$.}\label{fig5}
\end{figure}

In Fig. \ref{fig5}, we illustrate the achievable NEE performance with respect to the minimum data rate requirement ${R_{th}}$ of SU when ${P_{\max }}=25$ dBm. We can see that the achievable NEE decreases as ${R_{th}}$ increases for all the designs since the optimization feasible set becomes smaller. Specifically, the achievable NEE of our proposed NEE maximization design is almost unchanged when ${R_{th}}$ is sufficiently small, decreases rapidly when ${R_{th}}$ is relatively large, and finally approaches the value equal to that of WSR maximization design. This is attributable to the fact that more power is utilized in the NEE maximization design as ${R_{th}}$ increases, which even equals to the power required by the WSR maximization design, especially when ${R_{th}}$ is relatively large. It is also observed that our proposed path-following algorithm can achieve better performance than the Dinkelbach-ICA method for both NNPD and NPD cases. We finally notice that the performance gap between NPD and NNPD gradually decreases as ${R_{th}}$ increases. The reason is because, with high ${R_{th}}$, the optimization design for both cases are
taking more considerations to satisfy SU's rate requirement and achieve similar NEE performance finally.

\begin{figure}
  \centering
  \includegraphics[width=3.2in]{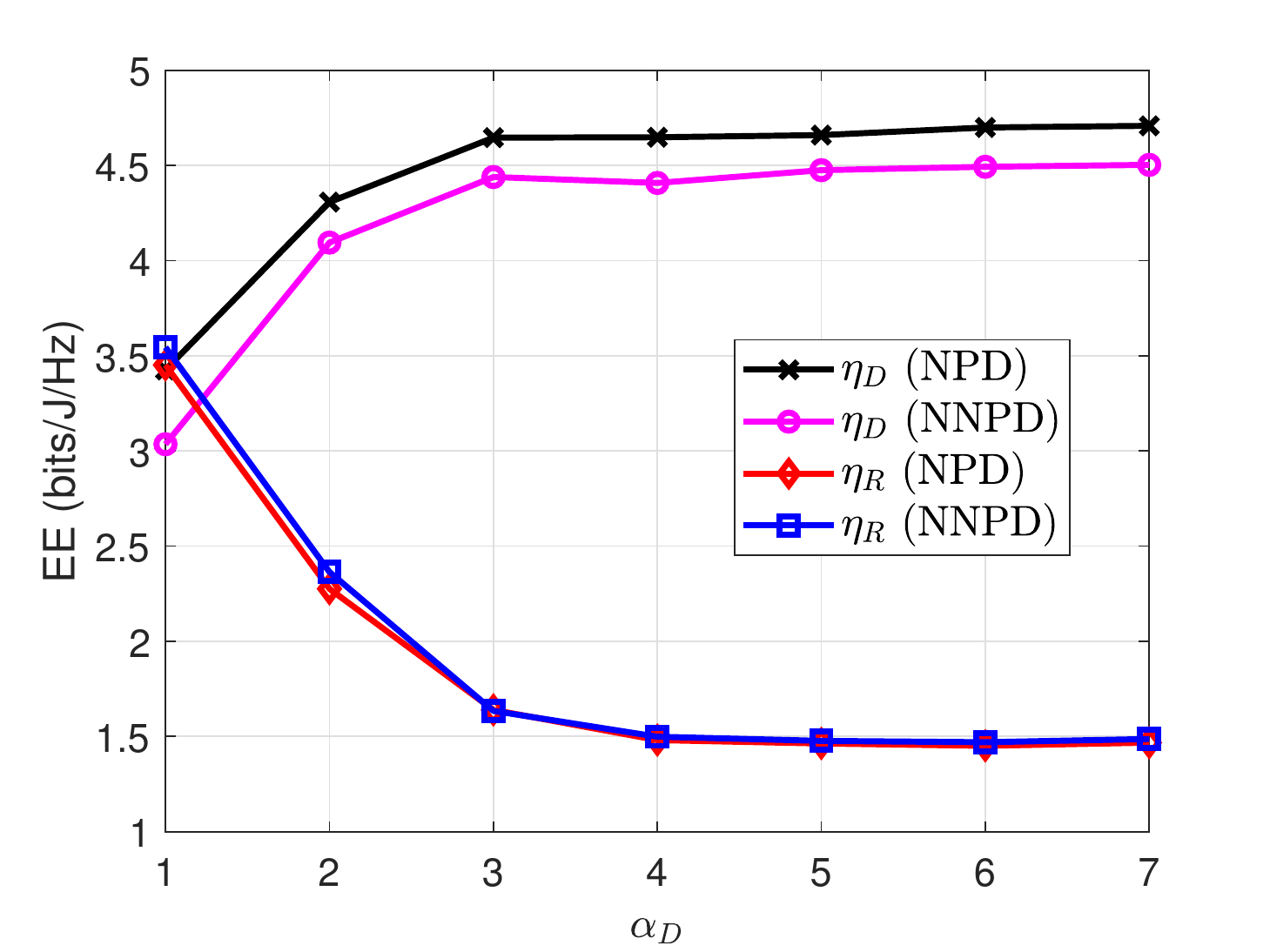}\\
\caption{Achievable EE versus the weight ${\alpha _D}$.}\label{fig6}
\end{figure}

Fig. \ref{fig6} shows the effect of weight ${\alpha _D}$ on the achievable EE performance obtained by our scheme when ${P_{\max }}=25$ dBm and ${\alpha _R}=1$. As ${\alpha _D}$ increases, the effective eavesdropping EE $\eta_D$ first gradually increases and finally saturates to a constant value, while the achievable EE of SU $\eta_R$ first decreases and then remains unchanged. This further demonstrates that the higher the weight ${\alpha _D}$, the more importance of eavesdropping performance, and vice versa, i.e., the higher the weight ${\alpha _R}$, the more importance of SU's performance (we omit the results here for brevity).
Furthermore, it is worth noting that NPD and NNPD achieve similar performance for $\eta_R$, while NPD outperforms NNPD for $\eta_D$ since a higher eavesdropping rate can be obtained by NPD.

\begin{figure}
  \centering
  \includegraphics[width=3.2in]{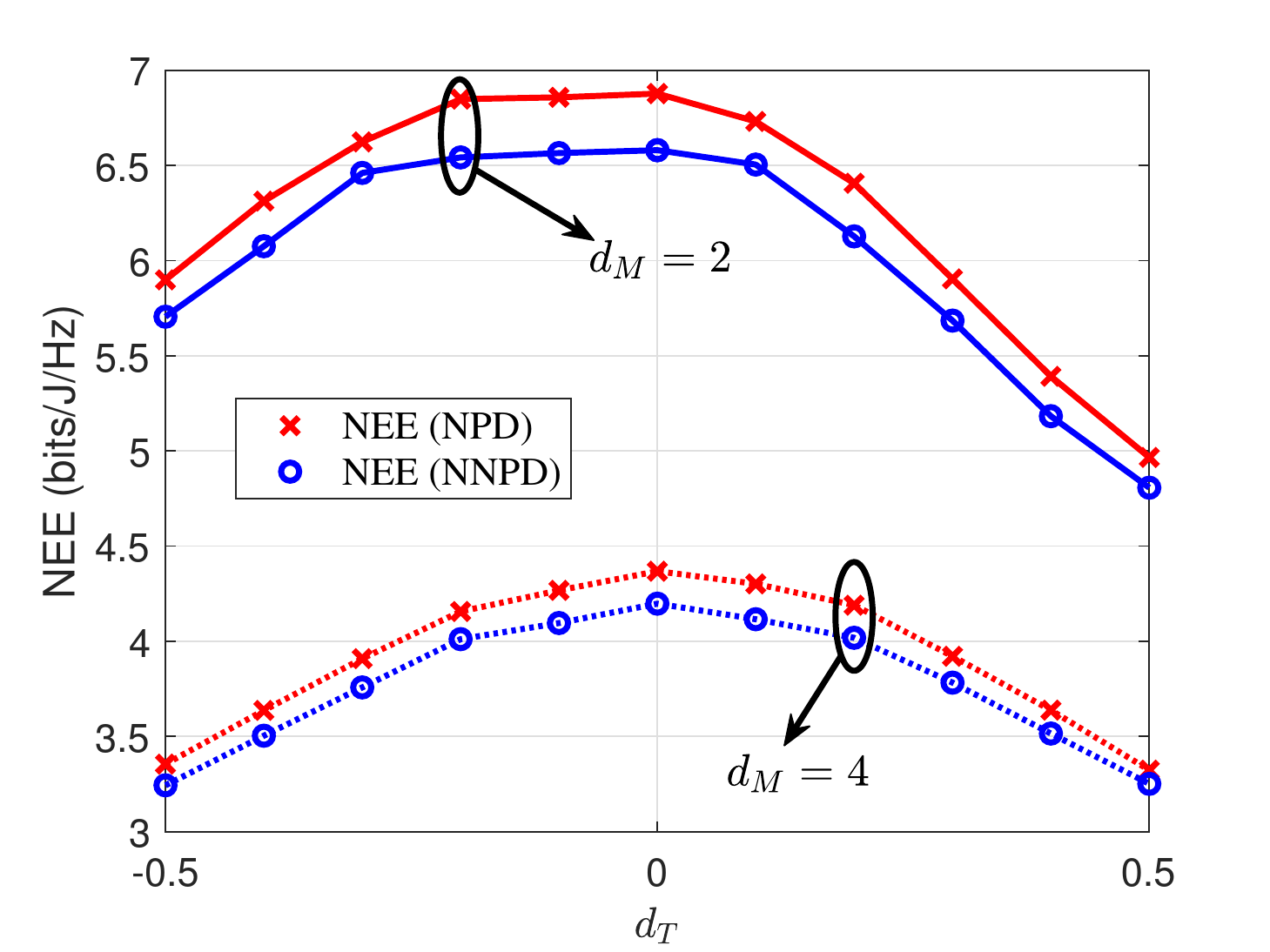}\\
\caption{Achievable NEE with different locations of T and M.}\label{fig7}
\end{figure}
Finally, we test the effect of network topology on the achievable NEE performance of our scheme in Fig. \ref{fig7}. Specifically, we compare the achievable NEE performance with different locations of T and M, where T and M are located in $({d_T},1)$ and $(0,{d_M})$, respectively. From Fig. \ref{fig7}, we observe that when M moves away from T, the achievable NEE monotonically decreases since the eavesdropping ability degrades when the distance between T and M increases, which results in a lower NEE. In addition, to perform relaying and jamming effectively, an optimum position for T can be found to obtain the optimal achievable NEE performance. These analyses imply that network topology is of great importance for achievable NEE performance. We again notice that the achieved NEE performance of NPD outperforms that achieved by NNPD for different locations of T and M, which further demonstrates the advantage of NPD over NNPD.


\begin{figure}
\begin{subfigure}{.5\textwidth}
  \centering
  \includegraphics[width=0.93\linewidth]{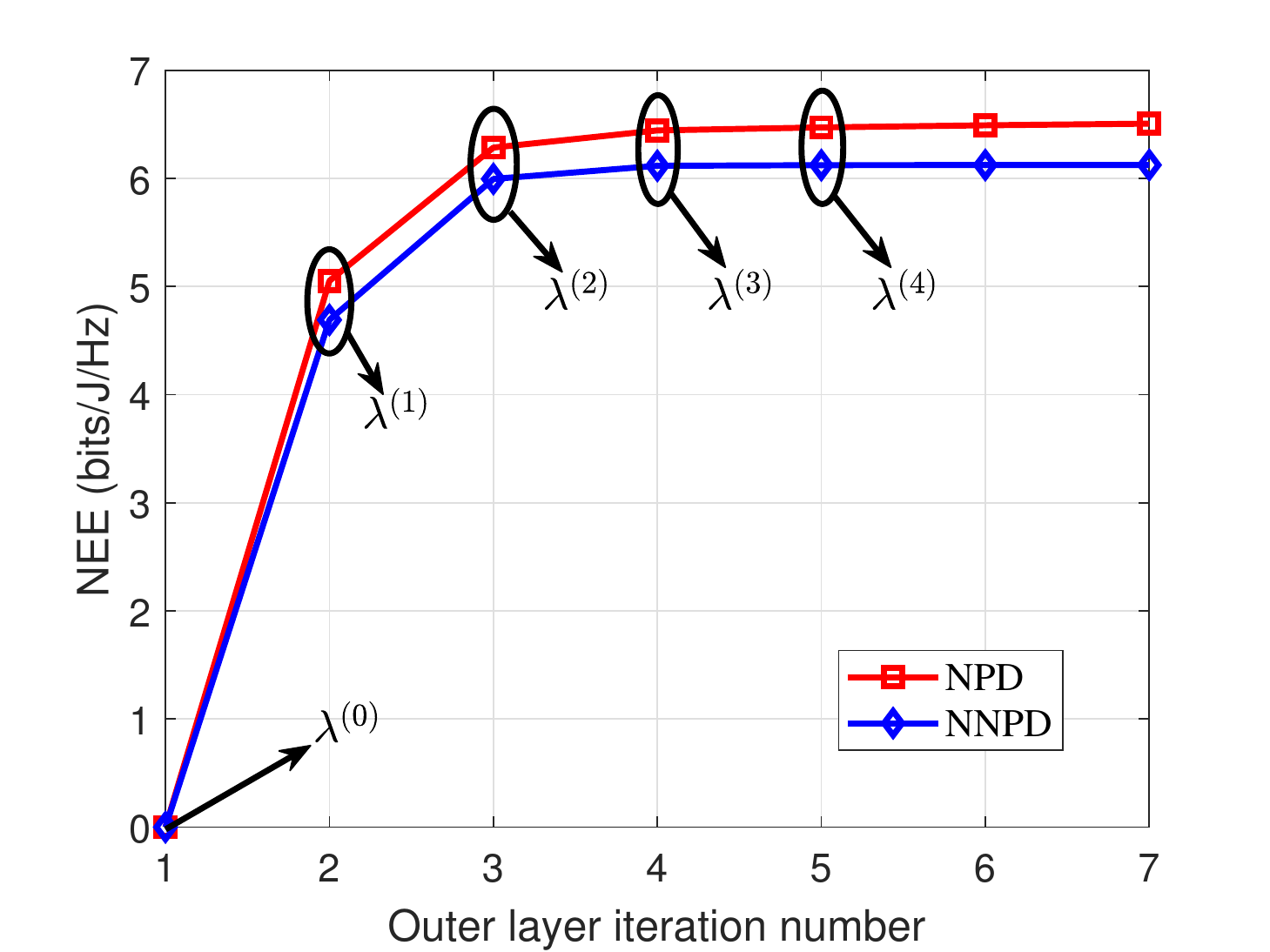}
  \caption{Convergence behaviour of outer layer optimization.}
  \label{AO_outer}
\end{subfigure}
\begin{subfigure}{.5\textwidth}
  \centering
  \includegraphics[width=0.93\linewidth]{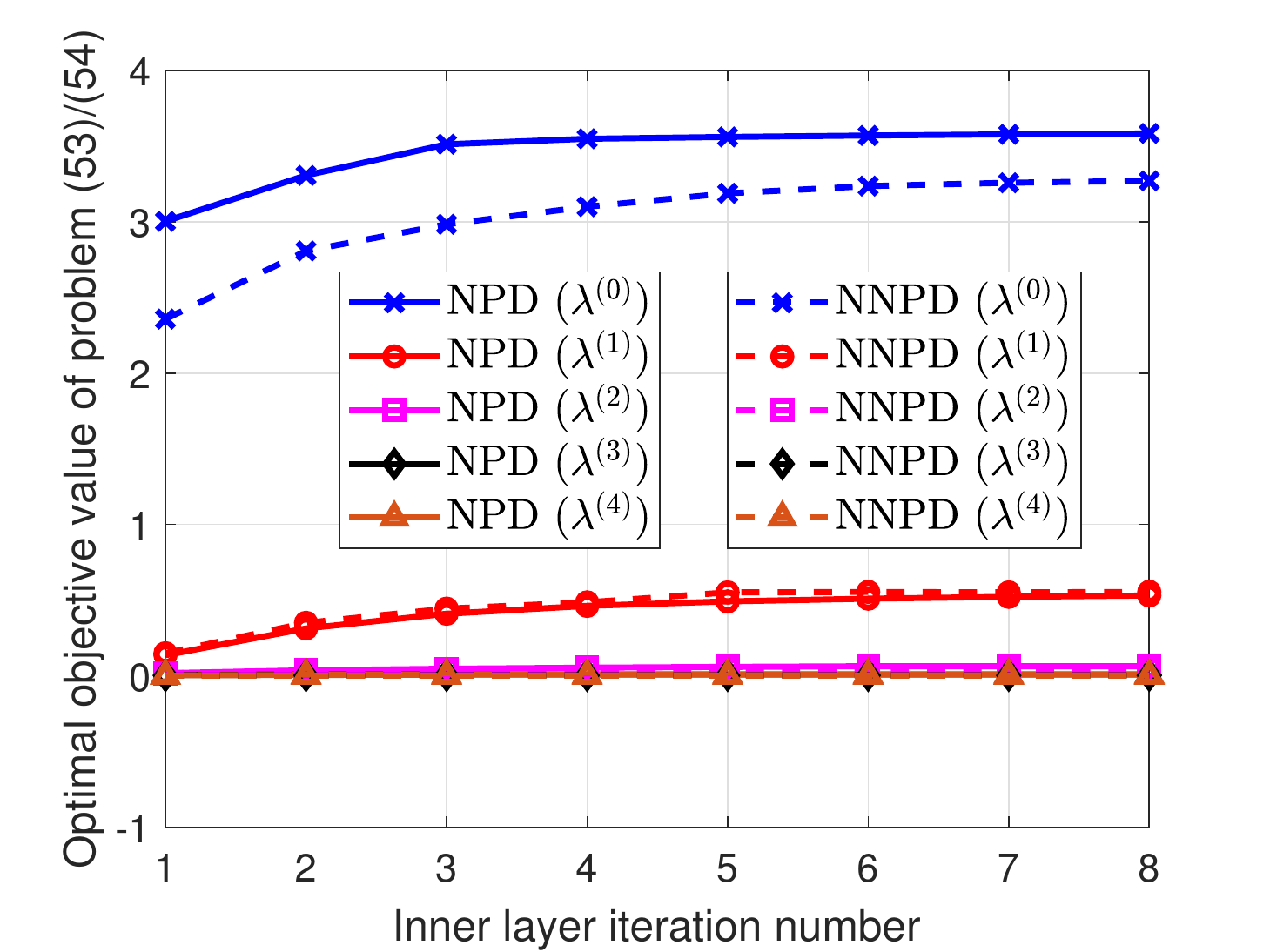}
  \caption{Convergence behaviour of inner layer optimization.}
  \label{AO_inner}
\end{subfigure}
\caption{Convergence behaviours of AO method.}
\label{AO_convergence}
\end{figure}
\subsection{Performance Evaluation of the Proposed Scheme
for Imperfect CSI}
In this subsection, we consider the imperfect CSI scenario and compare the performance of our proposed robust energy efficient design with that of perfect CSI results. To highlight superiority of the proposed scheme, we also compare the system outage probability performance with that of non-robust design, where the estimated channels are regarded as the actual channels and a solution can be obtained by our path-following algorithm in Section \ref{path_following}. For simplicity, we assume that ${\epsilon _{DS}}=\epsilon \left\| {{\hat h}_{DS}} \right\|$, ${\epsilon _{TS}}=\epsilon \left\| {{{\bf{\hat h}}}_{TS}} \right\|$, ${\epsilon _{DT}}=\epsilon \left\| {{{\bf{\hat h}}}_{DT}} \right\|$ and ${\epsilon _{RS}}=\epsilon \left\| {{\hat h}_{RS}} \right\|$.

Fig. \ref{AO_convergence} presents convergence behaviours of the proposed AO method for both outer layer optimization (Fig. \ref{AO_outer}) and inner layer optimization (Fig. \ref{AO_inner}) with $\epsilon=0.02$ and ${P_{\max }} = 25$ dBm. From Fig. \ref{AO_outer}, we observe that after four iterations, the outer layer optimization of AO method achieves a steady value for both NPD and NNPD scenarios, exhibiting a good convergence rate. The results in Fig. \ref{AO_inner} demonstrate that the inner optimization of AO method converges for each given $\lambda $, and as $\lambda $ increases, the optimal objective value of problem (\ref{Robust_NNPD})/(\ref{Robust_NPD}) decreases and finally approaches 0. These analyses strongly support the fast convergence property of our proposed AO method.

\begin{figure}
  \centering
  \includegraphics[width=3.2in]{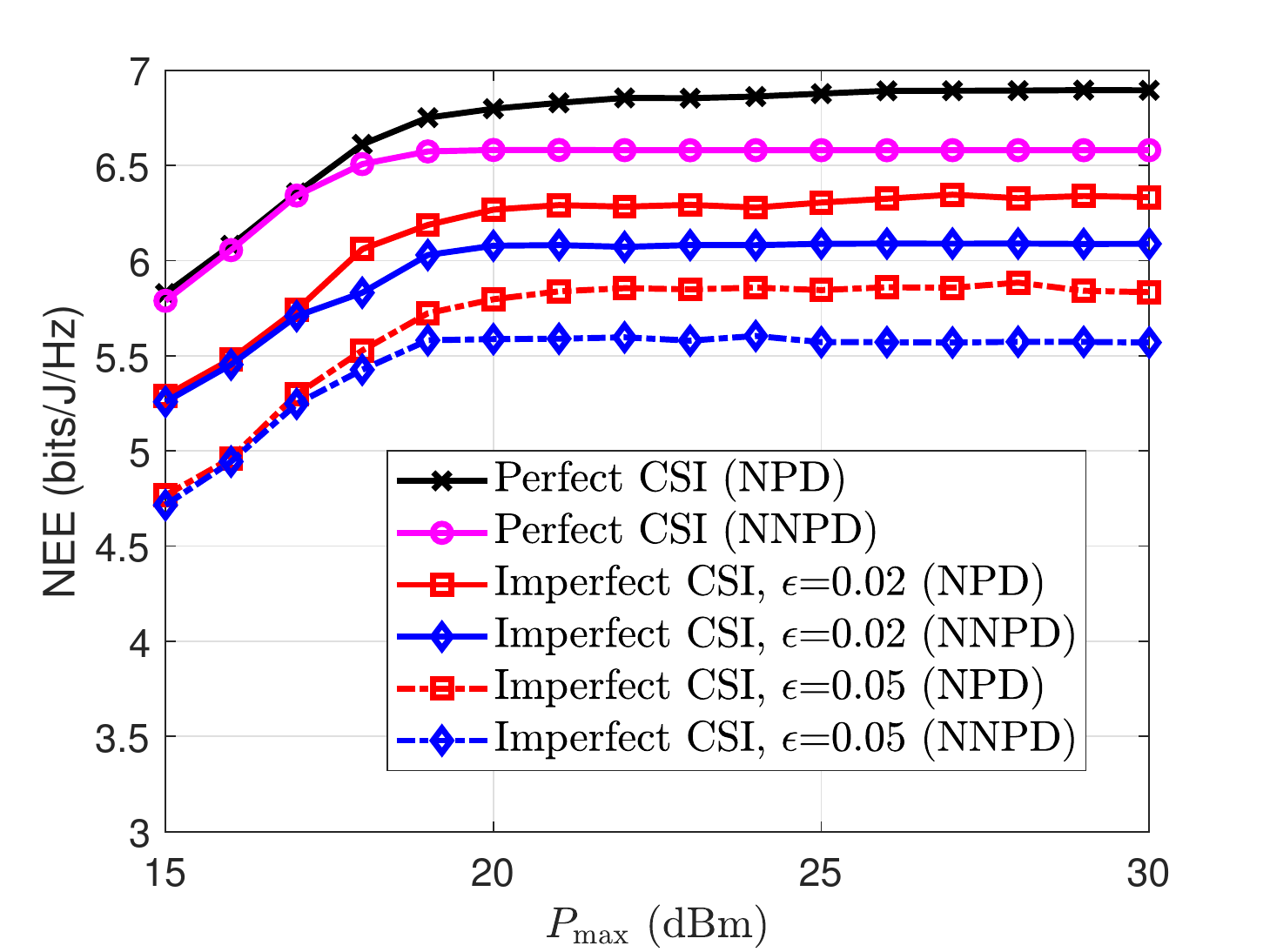}\\
\caption{Achievable NEE comparison for perfect and imperfect CSI cases.}\label{fig9_robust}
\end{figure}

In Fig. \ref{fig9_robust}, we compare the achievable NEE performance of perfect and imperfect CSI cases. It can be observed that the perfect CSI case attains the best performance, as anticipated. We also notice that the achievable NEE performance of our AO method decreases as $\epsilon$ increases, which is because larger channel uncertainty requires more resources to satisfy system requirements and results in higher NEE performance loss. Indeed, the performance advantages of NPD over NNPD still exist for the imperfect CSI scenario.

\begin{figure}
  \centering
  \includegraphics[width=3.2in]{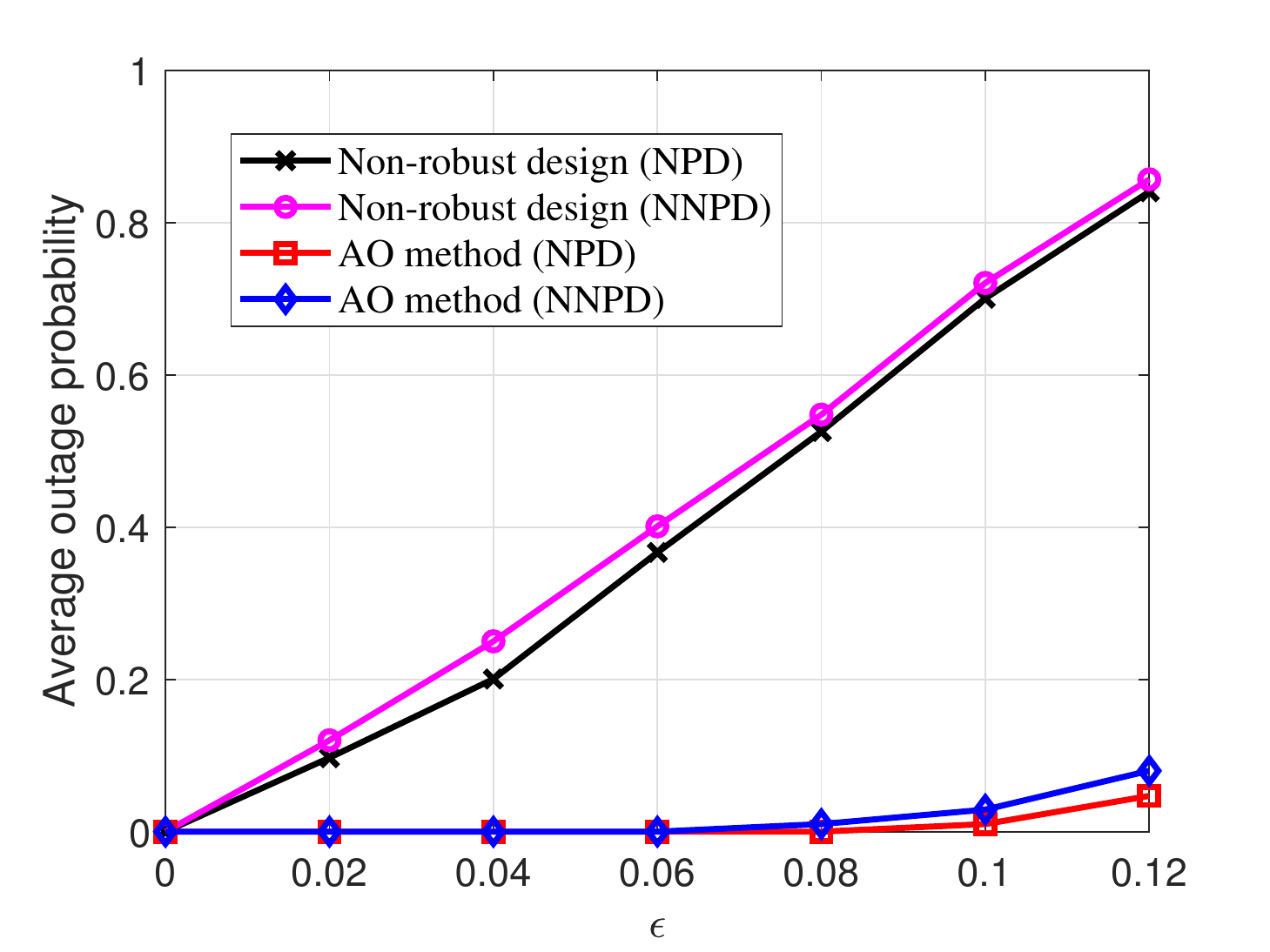}\\
\caption{Average outage probability versus channel uncertainty radius $\epsilon$.}\label{fig10_outage}
\end{figure}
Finally, we present the average system outage probability comparison for our AO method and non-robust design in Fig. \ref{fig10_outage} when ${R_{th}} = 1.5$ nats/s/Hz and ${P_{\max }} = 25$ dBm. If the achievable rate of the SU is lower than the minimum rate requirement ${R_{th}}$ or the rate of the eavesdropping link is smaller than that of the suspicious link, an outage occurs. From Fig. \ref{fig10_outage}, it is clear that the non-robust design is highly sensitive to CSI errors, i.e., even a slight increase of which may rapidly deteriorate the system outage probability performance. As $\epsilon$ increases, higher outage probability occurs for the non-robust design. However, our proposed AO method always outperforms the non-robust design and can efficiently satisfy the system requirements, although a slight outage probability occurs when $\epsilon$ is large. Furthermore, compared to NNPD, NPD achieves a smaller outage probability in high $\epsilon$ regions. These analyses demonstrate the robustness and effectiveness of our proposed AO method.

\section{Conclusion}\label{conclu}
In this paper, we studied a cooperative CR networks to facilitate distant proactive eavesdropping, where the secondary users help the monitor to eavesdrop on suspicious communication, and meanwhile, transmit its own information by sharing the spectrum belonging to the suspicious users. We considered both NNPD and NPD cases at the AF FD secondary transmitter. We jointly designed the AF relay matrix and precoding vector at secondary transmitter, as well as the receiver combining vector at monitor to maximize the NEE performance given the maximum available power at secondary transmitter and minimum data rate requirement of secondary user. We also guaranteed the successful eavesdropping condition, i.e., the achievable data rate of the eavesdropping link should be no less than that of the suspicious link. We then developed efficient path-following algorithm and robust AO method, respectively, for perfect and imperfect CSIs, to address the non-convexity of the formulated NEE maximization problem. We also analyzed the convergence and computational complexity of our proposed schemes. Numerical results validated the effectiveness of our schemes. It is demonstrated that the proposed schemes substantially outperform the benchmark schemes for both perfect and imperfect CSIs.


%

\appendix
For NNPD, by employing similar steps to those of constraint (\ref{Robust_MMSE_CR}), we can rewrite (\ref{Robust_MMSE_CD}), (\ref{Robust_MMSE_Ct}), and (\ref{Robust_MMSE_Cd}) as LMIs shown in (\ref{PD_susis_D}), (\ref{PD_ALTER}), and (\ref{PD_susis_Dd}), respectively. Here, we define
$
{{\bf{T}}_D} = \left[
 \begin{matrix}
   {{\beta _D} - {\lambda _{DS}} - {\lambda _{DT}} - {\lambda _{TS}}} & {\bm{\hat \phi} _D^H}  \\
   {{\bm{\hat \phi }_D}} & {\bf{I}}
  \end{matrix}
  \right]
$,
${{\bm{\theta }}_D} =  - {\left[ {{\epsilon _{DS}}\left[
 \begin{matrix}
   0 \\
   \bm{\Omega} _{DS}^D
  \end{matrix}
  \right],{\epsilon _{DT}}\left[
 \begin{matrix}
   {\bf{0}} \\
   \bm{\Omega} _{DT}^D
  \end{matrix}
  \right],{\epsilon _{TS}}\left[
 \begin{matrix}
   {\bf{0}} \\
   \bm{\Omega} _{TS}^D
  \end{matrix}
  \right]} \right]^H}$,
${\bm{\hat \phi }_D} = \left[
 \begin{matrix}
   {E_D}( {\sqrt {{P_S}} {D_D}{{\hat h}_{DS}} - 1}) \\
   \sqrt {{P_S}} {E_D}{D_D}{{{\bf{\hat h}}}_{DT}}{{\bf{V}}_{\bf{0}}}{\bf{G}}{{{\bf{\hat h}}}_{TS}} \\
   {\sigma _T}{E_D}{D_D}{( {{{{\bf{\hat h}}}_{DT}}{{\bf{V}}_{\bf{0}}}{\bf{G}}} )^H} \\
   {E_D}{D_D}{{{\bf{\hat h}}}_{DT}}{\bf{v}} \\
   {\sigma _D}{E_D}{D_D}
  \end{matrix}
  \right]$,
${E_D} = \sqrt {{S_D}}$,
$\bm{\Omega} _{DS}^D = {[\sqrt {{P_S}} {E_D}{D_D},0,{\bf{0}},0,0]^T}$,
$\bm{\Omega} _{DT}^D = \left[
 \begin{matrix}
   0 \\
   \sqrt {{P_S}} {E_D}{D_D}{( {{{\bf{V}}_{\bf{0}}}{\bf{G}}{{{\bf{\hat h}}}_{TS}}} )^H} \\
   {\sigma _T}{E_D}{D_D}{( {{{\bf{V}}_{\bf{0}}}{\bf{G}}} )^H} \\
   {E_D}{D_D}{{\bf{v}}^H} \\
   0
  \end{matrix}
  \right]$,
$\bm{\Omega} _{TS}^D = {[0,\sqrt {{P_S}} {E_D}{D_D}{{\bf{\hat h}}_{DT}}{{\bf{V}}_{\bf{0}}}{\bf{G}},{\bf{0}},0,0]^T}$,
${{\bf{T}}_t} = \left[
 \begin{matrix}
   {{\beta _t} - {\eta _{DT}} - {\eta _{TS}} - {\eta _{DS}}} & {\bm{\hat \phi} _t^H} \\
   {{\bm{\hat \phi }_t}} & {\bf{I}}
  \end{matrix}
  \right]$,
${{\bm{\theta }}_t} =  - {\left[ {{\epsilon _{DT}}\left[
 \begin{matrix}
   {\bf{0}}  \\
   \bm{\Omega} _{DT}^t
  \end{matrix}
  \right],{\epsilon _{TS}}\left[
 \begin{matrix}
   {\bf{0}}  \\
   \bm{\Omega} _{TS}^t
  \end{matrix}
  \right],{\epsilon _{DS}}\left[
 \begin{matrix}
   0  \\
   \bm{\Omega} _{DS}^t
  \end{matrix}
  \right]} \right]^H}$,
${\bm{\hat \phi }_t} = \left[
 \begin{matrix}
   {E_t}  \\
   \sqrt {{P_S}} \sigma _D^{ - 1}{E_t}{{{\bf{\hat h}}}_{DT}}{{\bf{V}}_{\bf{0}}}{\bf{G}}{{{\bf{\hat h}}}_{TS}} \\
   \sigma _D^{ - 1}{\sigma _T}{E_t}{( {{{{\bf{\hat h}}}_{DT}}{{\bf{V}}_{\bf{0}}}{\bf{G}}} )^H} \\
   \sigma _D^{ - 1}{E_t}{{{\bf{\hat h}}}_{DT}}{\bf{v}} \\
   \sigma _D^{ - 1}\sqrt {{P_S}} {E_t}{{\hat h}_{DS}}
  \end{matrix}
  \right]$,
${E_t} = \sqrt {{S_t}}$,
$\bm{\Omega} _{DT}^t = \left[
 \begin{matrix}
   0 \\
   \sqrt {{P_S}} \sigma _D^{ - 1}{E_t}{( {{{\bf{V}}_{\bf{0}}}{\bf{G}}{{{\bf{\hat h}}}_{TS}}} )^H} \\
   \sigma _D^{ - 1}{\sigma _T}{E_t}{( {{{\bf{V}}_{\bf{0}}}{\bf{G}}} )^H} \\
   \sigma _D^{ - 1}{E_t}{{\bf{v}}^H} \\
   0
  \end{matrix}
  \right]$,
$\bm{\Omega} _{TS}^t = {[0,\sqrt {{P_S}} \sigma _D^{ - 1}{E_t}{{\bf{\hat h}}_{DT}}{{\bf{V}}_{\bf{0}}}{\bf{G}},{\bf{0}},0,0]^T}$,
$\bm{\Omega} _{DS}^t = {[0,0,{\bf{0}},0,\sigma _D^{ - 1}\sqrt {{P_S}} {E_t}]^T}$,
${{\bm{\theta }}_d} =  - {\left[ {{\epsilon _{DT}}\left[
 \begin{matrix}
   {\bf{0}} \\
   \bm{\Omega} _{DT}^d
  \end{matrix}
  \right],{\epsilon _{TS}}\left[
 \begin{matrix}
   {\bf{0}} \\
   \bm{\Omega} _{TS}^d
  \end{matrix}
  \right]} \right]^H}$,
${\bm{\hat \phi }_d} = \left[
 \begin{matrix}
   {E_d}( {{{\bf{D}}_{\bf{d}}}{{[ {\sqrt {{P_S}} {{{\bf{\hat h}}}_{DT}}{{\bf{V}}_{\bf{0}}}{\bf{G}}{{{\bf{\hat h}}}_{TS}},{\sigma _T}{{{\bf{\hat h}}}_{DT}}{{\bf{V}}_{\bf{0}}}{\bf{G}},{{{\bf{\hat h}}}_{DT}}{\bf{v}}} ]}^H} - 1} ) \\
   {\sigma _D}{E_d}{\bf{D}}_{\bf{d}}^H
  \end{matrix}
  \right]$,
${E_d} = \sqrt {{S_d}}$,
$\bm{\Omega} _{DT}^d = \left[
 \begin{matrix}
   {E_d}{{\bf{D}}_{\bf{d}}}{[ {\sqrt {{P_S}} {{\bf{V}}_{\bf{0}}}{\bf{G}}{{{\bf{\hat h}}}_{TS}},{\sigma _T}{{\bf{V}}_{\bf{0}}}{\bf{G}},{\bf{v}}}]^H} \\
   {\bf{0}}
  \end{matrix}
  \right]$ and
$\bm{\Omega} _{TS}^d = \left[
 \begin{matrix}
   {E_d}{{\bf{D}}_{\bf{d}}}{[ {\sqrt {{P_S}} {{( {{{{\bf{\hat h}}}_{DT}}{{\bf{V}}_{\bf{0}}}{\bf{G}}} )}^H},{\bf{0}},{\bf{0}}} ]^H} \\
   {\bf{0}}
  \end{matrix}
  \right]$.

Similarly for NPD, we can have LMIs shown in (\ref{PD_susis_D_NPD}), (\ref{PD_ALTER_NPD}), and (\ref{PD_susis_Dd_NPD}) to replace constraints (\ref{Robust_MMSE_CD}), (\ref{Robust_MMSE_Ct}), and (\ref{Robust_MMSE_Cd}), respectively. Here, we denote
${{{\bf{\bar T}}}_D} = \left[
 \begin{matrix}
   {{\beta _D} - {\lambda _{DS}} - {\lambda _{DT}} - {\lambda _{TS}}} & {\hat {\bar {\bm{\phi}}} _D^H} \\
   \hat{ \bar{ \bm{\phi}}}_D & {\bf{I}}
  \end{matrix}
  \right]$,
${{{\bm{\bar \theta }}}_D} =  - {\left[ {{\epsilon _{DS}}\left[
 \begin{matrix}
   0 \\
   \bm{\bar \Omega} _{DS}^D
  \end{matrix}
  \right],{\epsilon _{DT}}\left[
 \begin{matrix}
   {\bf{0}} \\
   \bm{\bar \Omega} _{DT}^D
  \end{matrix}
  \right],{\epsilon _{TS}}\left[
 \begin{matrix}
   {\bf{0}} \\
   \bm{\bar \Omega} _{TS}^D
  \end{matrix}
  \right]} \right]^H}$,
$ \hat{ \bar{ \bm{\phi}}}_D =  \left[
 \begin{matrix}
   {E_D}[ {\sqrt {{P_S}} {D_D}( {{{\hat h}_{DS}} + {{{\bf{\hat h}}}_{DT}}{{\bf{V}}_{\bf{0}}}{\bf{G}}{{{\bf{\hat h}}}_{TS}}} ) - 1} ] \\
   {\sigma _T}{E_D}{D_D}{( {{{{\bf{\hat h}}}_{DT}}{{\bf{V}}_{\bf{0}}}{\bf{G}}} )^H} \\
   {E_D}{D_D}{{{\bf{\hat h}}}_{DT}}{\bf{v}} \\
   {\sigma _D}{E_D}{D_D}
  \end{matrix}
  \right]$,
${E_D} = \sqrt {{S_D}} $,
$\bm{\bar \Omega} _{DS}^D = {[\sqrt {{P_S}} {E_D}{D_D},{\bf{0}},0,0]^T}$,
$\bm{\bar \Omega} _{DT}^D = \left[
 \begin{matrix}
   \sqrt {{P_S}} {E_D}{D_D}{( {{{\bf{V}}_{\bf{0}}}{\bf{G}}{{{\bf{\hat h}}}_{TS}}} )^H} \\
   {\sigma _T}{E_D}{D_D}{( {{{\bf{V}}_{\bf{0}}}{\bf{G}}})^H} \\
   {E_D}{D_D}{{\bf{v}}^H} \\
   0
  \end{matrix}
  \right]$,
$\bm{\bar \Omega} _{TS}^D = {[\sqrt {{P_S}} {E_D}{D_D}{{\bf{\hat h}}_{DT}}{{\bf{V}}_{\bf{0}}}{\bf{G}},{\bf{0}},0,0]^T}$,
${{{\bf{\bar T}}}_t} = \left[
 \begin{matrix}
   {{\beta _t} - {\eta _{DT}} - {\eta _{TS}} - {\eta _{DS}}} & {\hat {\bar {\bm{\phi}}} _t^H} \\
   \hat{ \bar{ \bm{\phi}}}_t & {\bf{I}}
  \end{matrix}
  \right]$,
${{{\bm{\bar \theta }}}_t} =  - {\left[ {{\epsilon _{DT}}\left[
 \begin{matrix}
   {\bf{0}} \\
   \bm{\bar \Omega} _{DT}^t
  \end{matrix}
  \right],{\epsilon _{TS}}\left[
 \begin{matrix}
   {\bf{0}} \\
   \bm{\bar \Omega} _{TS}^t
  \end{matrix}
  \right],{\epsilon _{DS}}\left[
 \begin{matrix}
   0 \\
   \bm{\bar \Omega} _{DS}^t
  \end{matrix}
  \right]} \right]^H}$,
$\hat{ \bar{ \bm{\phi}}}_t = \left[
 \begin{matrix}
   {E_t} \\
   \sqrt {{P_S}} \sigma _D^{ - 1}{E_t}( {{{\hat h}_{DS}} + {{{\bf{\hat h}}}_{DT}}{{\bf{V}}_{\bf{0}}}{\bf{G}}{{{\bf{\hat h}}}_{TS}}} ) \\
   \sigma _D^{ - 1}{\sigma _T}{E_t}{( {{{{\bf{\hat h}}}_{DT}}{{\bf{V}}_{\bf{0}}}{\bf{G}}} )^H} \\
   \sigma _D^{ - 1}{E_t}{{{\bf{\hat h}}}_{DT}}{\bf{v}}
  \end{matrix}
  \right]$,
${E_t} = \sqrt {{S_t}}$,
$\bm{\bar \Omega} _{DT}^t = \left[
 \begin{matrix}
   0 \\
   \sqrt {{P_S}} \sigma _D^{ - 1}{E_t}{( {{{\bf{V}}_{\bf{0}}}{\bf{G}}{{{\bf{\hat h}}}_{TS}}} )^H} \\
   \sigma _D^{ - 1}{\sigma _T}{E_t}{( {{{\bf{V}}_{\bf{0}}}{\bf{G}}})^H} \\
   \sigma _D^{ - 1}{E_t}{{\bf{v}}^H}
  \end{matrix}
  \right]$,
$\bm{\bar \Omega} _{TS}^t = {[0,\sqrt {{P_S}} \sigma _D^{ - 1}{E_t}{{\bf{\hat h}}_{DT}}{{\bf{V}}_{\bf{0}}}{\bf{G}},{\bf{0}},0]^T}$,
$\bm{\bar \Omega} _{DS}^t = {[0,\sqrt {{P_S}} \sigma _D^{ - 1}{E_t},{\bf{0}},0]^T}$,
${{{\bm{\bar \theta }}}_d} =  - {\epsilon _{DT}}[ {{\bf{0}}\;{{( {\bm{\bar \Omega }_{DT}^d} )}^H}} ]$,
$\hat{\bar{\bm{\phi}}}_d = \left[
 \begin{matrix}
   {E_d}( {{{\bf{D}}_{\bf{d}}}{{[ {{\sigma _T}{{{\bf{\hat h}}}_{DT}}{{\bf{V}}_{\bf{0}}}{\bf{G}},{{{\bf{\hat h}}}_{DT}}{\bf{v}}}]}^H} - 1} ) \\
   {\sigma _D}{E_d}{\bf{D}}_{\bf{d}}^H
  \end{matrix}
  \right]$,
${E_d} = \sqrt {{S_d}}$ and
$\bm{\bar \Omega} _{DT}^d = \left[
 \begin{matrix}
   {E_d}{{\bf{D}}_{\bf{d}}}{\left[ {{\sigma _T}{{\bf{V}}_{\bf{0}}}{\bf{G}},{\bf{v}}} \right]^H} \\
   {\bf{0}}
  \end{matrix}
  \right]$.


%
%

\ifCLASSOPTIONcaptionsoff
  \newpage
\fi



%




\bibliographystyle{IEEEtran}
\footnotesize
\bibliography{ref_secrecy}

%

\begin{IEEEbiography}[{\includegraphics[width=1in,height=1.25in,clip,keepaspectratio]{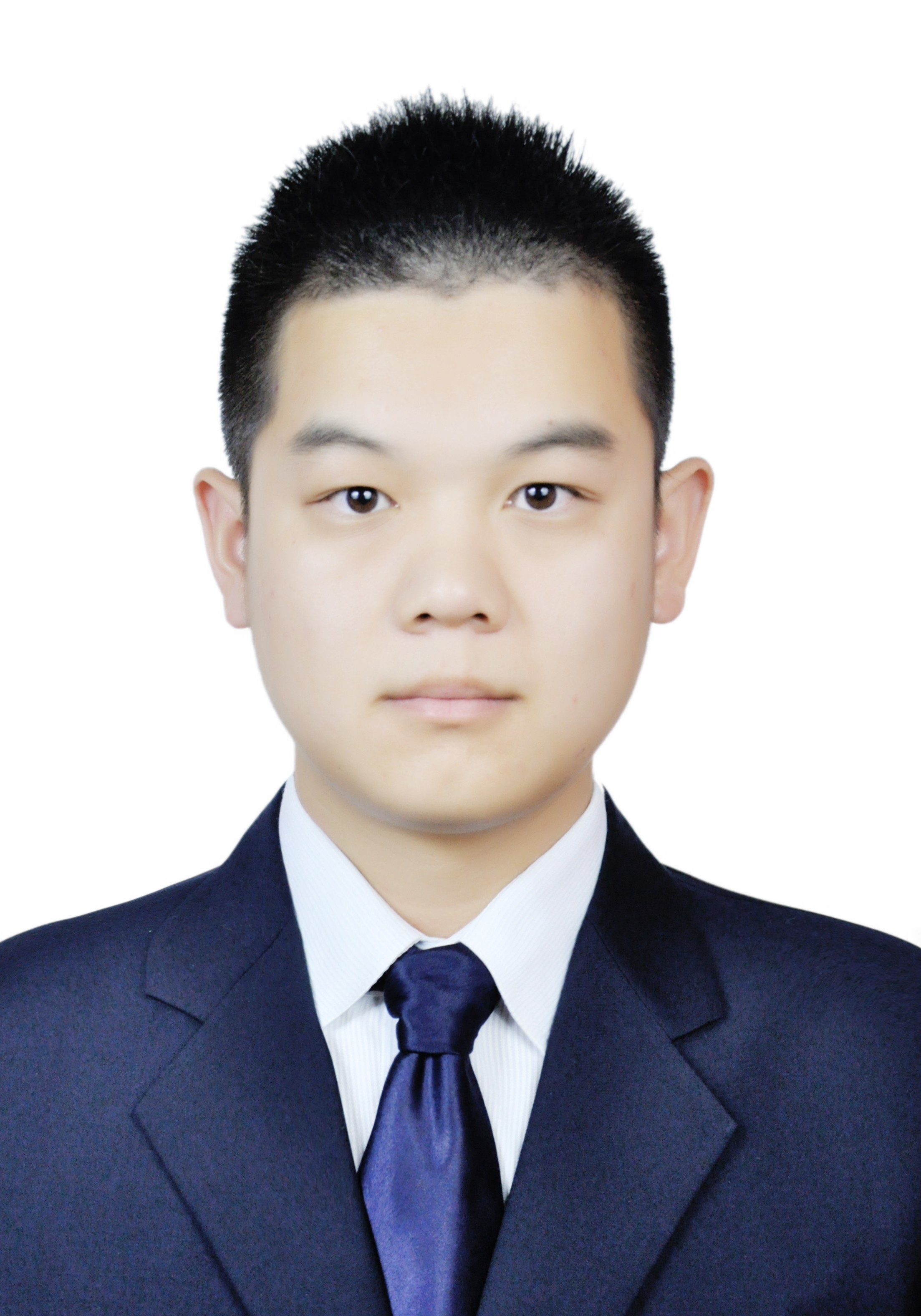}}]{Yao Ge}(Member, IEEE) received the B.Eng. degree in Electronics and Information Engineering and the M.Eng. degree (research) in Communication and Information System from Northwestern Polytechnical University (NPU), Xi’an, China, in 2013 and 2016, respectively, and the Ph.D. degree in Electronic Engineering from The Chinese University of Hong Kong (CUHK), Shatin, Hong Kong, in 2021.
He is currently a Research Fellow with Continental-NTU Corporate Lab, Nanyang Technological University (NTU), Singapore.
From October 2015 to March 2016, he was a Visiting Student with the Department of Electrical and Computer Systems Engineering, Monash University, Melbourne, VIC, Australia. From April 2016 to August 2016, he was a Visiting Student with the Department of Computer, Electrical and Mathematical Science \& Engineering, King Abdullah University of Science and Technology (KAUST), Thuwal, Saudi Arabia. From May 2019 to December 2019, he was a Visiting Student with the Department of Electrical and Computer Engineering, University of California at Davis (UC Davis), Davis, CA, USA.
He is the Founding Member of the IEEE ComSoc special interest
group (SIG) on OTFS.
His current research interests include wireless communications, Internet of Things, cognitive radio networks, automotive vehicle signal processing and communications, wireless network security, and statistical signal processing.
\end{IEEEbiography}

\begin{IEEEbiography}[{\includegraphics[width=1in,height=1.25in,clip,keepaspectratio]{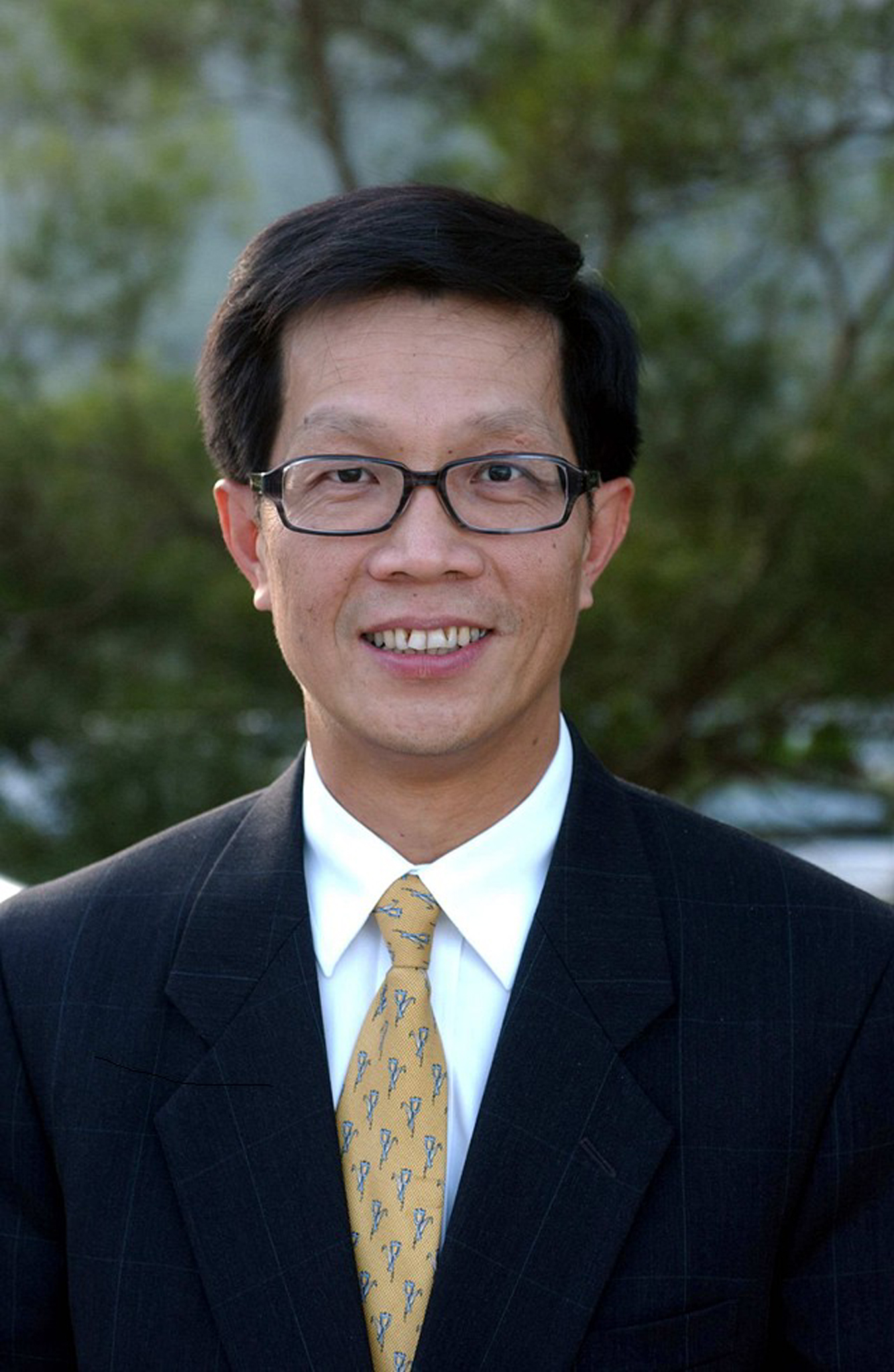}}]{P. C. Ching}(Fellow, IEEE) received the B.Eng. (Hons.) and Ph.D. degrees from the University of Liverpool, U.K., in 1977 and 1981, respectively. From 1981 to 1982, he was a Research Officer with the University of Bath, U.K. In 1982, he returned to Hong Kong and then joined the Hong Kong Polytechnic as a Lecturer. Since 1984, he has been with the Department of Electronic Engineering, The Chinese University of Hong Kong (CUHK), where he
is currently a Choh-Ming Li Professor of Electronic Engineering. He was the Department Chairman from 1995 to 1997, the Dean of Engineering from 1997 to 2003, and the Head of Shaw College from 2004 to 2008. He became the Director of the Shun Hing Institute of Advanced Engineering in 2004. From 2006 to 2014, Prof. Ching was appointed as a Pro-Vice-Chancellor/Vice-President of CUHK. From 2013 to 2014, he also took up the Directorship of the CUHK Shenzhen Research Institute. He is very active in promoting professional activities, both in Hong Kong and overseas. He was a Council Member of the Institution of
Electrical Engineers (IEE), a Past Chairman of the IEEE Hong Kong Section, an Associate Editor of {\sc IEEE Transactions on Signal Processing} from 1997 to 2000 and {\sc IEEE Signal Processing Letters} from 2001 to 2003. He was also a member of the Technical Committee of the IEEE Signal Processing Society from 1996 to 2004. He was appointed as the Editor-in-Chief of the {\em HKIE Transactions} from 2001 to 2004. He has been an Honorary Member of the Editorial Committee for the {\em Journal of Data Acquisition and Processing} since 2000. He has been instrumental in organizing many international conferences in Hong Kong including the 1997 IEEE International Symposium on Circuits and Systems, where he was the Vice-Chairman. He
also served as a Technical Program Co-Chair for the 2003 and 2016 IEEE International Conference on Acoustics, Speech and Signal Processing. He received the IEEE Third Millennium Award in 2000 and the HKIE Hall of Fame in 2010. In addition, he also plays an active role in community services. He received the Silver Bauhinia Star (SBS) and the Bronze Bauhinia
Star (BBS) by the HKSAR Government in 2017 and 2010, respectively, in recognition of his long and distinguished public and community services. He is currently the Chairman of the Board of Directors of the Nano and Advanced Materials Institute Limited, a Council Member of the Shaw Prize Foundation, and a member of the Museum Advisory Committee (MAC) and the Chairperson of its Science Sub-committee. He is elected as the President of the Hong Kong Academy of Engineering Sciences (HKAES) in 2018.
\end{IEEEbiography}




\end{document}